\documentclass[fleqn,usenatbib]{mnras}

\usepackage{newtxtext,newtxmath}

\usepackage[T1]{fontenc}
\usepackage{ae,aecompl}
\usepackage{times}

\usepackage{graphicx}	
\usepackage{amsmath}	
\usepackage{gensymb}
\usepackage{subfig}

\usepackage{hyperref}
\usepackage{enumitem}

\newcommand{\msun}{{\,\rm M_\odot}}

\title[Gravothermal Collapse in Velocity Dependent Self-Interacting Dark Matter]{The Onset of Gravothermal Core Collapse in Velocity Dependent Self-Interacting Dark Matter Subhaloes}

\author[H. C. Turner et al.]{
Hannah C. Turner,$^{1,2}$\thanks{E-mail: hannah.c.turner@durham.ac.uk}
Mark R. Lovell,$^{3, 4}$
Jes{\'u}s Zavala$^{3}$ and Mark Vogelsberger$^{5}$
\\
$^{1}$Centre for Extragalactic Astronomy, Durham University, South Road, Durham DH1 3LE, UK\\
$^{2}$Department of Physics and Mathematics, University of Hull,
HU6 7RX\\ 
\linebreak
$^{3}$Center for Astrophysics and Cosmology, Science Institute,
University of Iceland, Dunhagi 5, 107 Reykjavik, Iceland\\ \linebreak
$^{4}$Institute for Computational Cosmology, Durham University, South Road, Durham DH1 3LE, UK\\
$^{5}$Department of Physics, Kavli Institute for Astrophysics and Space Research, Massachusetts Institute of Technology, Cambridge, MA 02139, USA}

\date{Accepted 2021 June 29. Received 2021 May 18; in original form 2020 October 27}

\pubyear{2020}

\begin{document}
\label{firstpage}
\pagerange{\pageref{firstpage}--\pageref{lastpage}}
\maketitle

\begin{abstract}
It has been proposed that gravothermal collapse due to dark matter self-interactions (i.e. self-interacting dark matter, SIDM) can explain the observed diversity of the Milky Way (MW) satellites' central dynamical masses. We investigate the process behind this hypothesis using an $N$-body simulation of a MW-analogue halo with velocity dependent self-interacting dark matter (vdSIDM) in which the low velocity self-scattering cross-section, $\sigma_\rmn{T}/m_\rmn{x}$, reaches 100~cm$^{2}$\,g$^{-1}$; we dub this model the vd100 model. We compare the results of this simulation to simulations of the same halo that employ different dark models, including cold dark matter (CDM) and other, less extreme SIDM models. The masses of the vd100 haloes are very similar to their CDM counterparts, but the values of their maximum circular velocities, $V_\rmn{max}$, are significantly higher. We determine that these high $V_\rmn{max}$ subhaloes were objects in the mass range  [$5\times10^{6}$, $1\times10^{8}$]~$\msun$ at $z=1$ that undergo gravothermal core collapse. These collapsed haloes have density profiles that are described by single power laws down to the resolution limit of the simulation, and the inner slope of this density profile is approximately $-3$. Resolving the ever decreasing collapsed region is challenging, and tailored simulations will be required to model the runaway instability accurately at scales $<1$~kpc.
\end{abstract}

\begin{keywords}
dark matter -- galaxies: haloes
\end{keywords}



\section{Introduction}
Self-interacting dark matter (SIDM) has been proposed to resolve conflicts between observations and the predictions of cosmological $N$-body simulations under the collisionless, cold dark matter (CDM) model. Despite the CDM model's significant achievements in explaining the distribution of matter on scales $>1$~Mpc \citep{wmap1,Eisenstein05,Planck16}, it falls short in its ability to explain observations of smaller scale structures, and in particular the distribution of mass in Local Group dwarf galaxies. We discuss these challenges in detail below; for a general review see \citet{bullock2017}.

 First, one would expect the central masses of the observed Milky Way (MW) satellites to be consistent with those of the most massive CDM subhaloes in $N$-body simulations. However, \citet{BoylanKolchin11,BoylanKolchin12} compared CDM subhaloes from the Aquarius simulation suite \citep{Springel08b} with the observed masses of MW satellite galaxies \citep{Walker09,Wolf10} and reported that the subhaloes are too centrally dense under the CDM model to host the luminous MW satellites. They showed that each of the Aquarius MW-halo analogue simulations hosted at least 4 subhaloes with a central density 2$\sigma$ higher than every observed MW satellite galaxy; this tension has become known as the `too-big-to-fail' (TBTF) problem.

A further small scale issue encountered by the CDM model is the cusp versus core problem, plus the apparent diversity in the central density of MW dwarf satellites. Whilst the CDM model predicts haloes with an inner density profile that is well described as a power law with a slope of $-1$ \citep{NFW_96, NFW_97, Moore98, Klypin01}, or the marginally shallower Einasto profile \citep{Einasto65} in the case of subhaloes \citep{Springel08b}, observations provide evidence for a significant number of haloes exhibiting cored density profiles with a constant central density \citep[see][]{Spergel00, Dave01, Gilmore07, Walker11}, although this interpretation is disputed \citep{Strigari10,Strigari17}, and could instead be an expression of non-spherical symmetry in these galaxies \citep{Hayashi20}. Recent studies into the diversity of satellite densities \citep{Kamada17, Zavala2019} and isolated dwarf rotation curves \citep{Oman15,Oman19,Santos19} for fixed halo properties show an open issue in relation to the apparently diverse nature of dwarf galaxy rotation curves.

The SIDM model has been shown to have the potential to resolve these issues both with $N$-body and full hydrodynamic simulations \citep[e.g.][]{Dave01, Colin02, Vogelsberger12, Rocha13, Zavala13, Fitts19, Vogelsberger2019}. The introduction of self-interactions with a cross section of order 1~cm$^{2}$\,g$^{-1}$ creates cores in MW satellite subhaloes, thus solving the core versus cusp problem directly and at the same time decreasing the central densities sufficiently to alleviate the TBTF tension. Many of these studies found that a velocity dependence of the self-interaction cross-section, or velocity-dependent SIDM (vdSIDM) model \citep{Vogelsberger12}, was required in order to obtain cores in dwarf galaxies while simultaneously preventing MW-mass haloes from becoming more spherical than observations permit \citep[see][]{Vogelsberger12, Peter13, Zavala13}. 

The next challenge for such models was to ensure that they could match the densities of all MW satellites simultaneously. \citet{Zavala2019} argued that this was unlikely to be the case for the SIDM models that had been considered previously: models with large values of the cross-section at low velocities ($\sim3$~cm$^{2}$\,g$^{-1}$ ) over-suppressed the central densities in ultra-faint dwarfs, whereas lower cross-section would be indistinguishable from CDM and therefore suffer from the TBTF problem in the classical dwarfs. In summary, it was not possible for these models to match the densities of both the ultrafaint satellites and the classical satellites simultaneously.

One proposed solution to this problem is the counter intuitive suggestion to make the cross-section much {\it larger} than $\sim3$~cm$^{2}$\,g$^{-1}$. If the cross-section is $>10$~cm$^{2}$\,g$^{-1}$ then the halo can undergo a process known as {\it gravothermal collapse}. The gravothermal collapse of an object, synonymous with the gravothermal catastrophe coined by \citet{LyndenBell68}, describes the way in which a system develops a negative heat capacity through the constant outwards transfer of energy. In the context of the MW satellite-analogue subhaloes orbiting a MW-size halo, this process proceeds as follows. The subhalo is initially cored, and since the majority of self-interactions occur within the central, core, most of the energy re-distribution (heat transfer), occurs there. The core radius increases proportionally with the inflow of heat until an isothermal core is formed that reaches the outer parts of the subhalo. This `colder' region has a negative temperature gradient so acts as a `thermal reservoir' and as such, triggers the onset of gravothermal core collapse as in-flowing heat drives the core to lower temperatures, preventing the realization of a thermal equilibrium \citep[see][]{Colin02, Glass10, Huo19}. The net result of this effect is to change the shape of the profile from cored to a very steep cusp.

Gravothermal core collapse has been cited to be a trigger for the formation of super massive black holes in the centre of dark matter haloes, without the need for the presence of baryons or a pre-existing black hole seed mechanism, if the core mass at the time of collapse is between 10$^{-8}$ and 10$^{-6}$ times that of the total halo mass \citep{Balberg02}. Furthermore, it has been proposed by \citet{Nishikawa20} \citep[see also][]{Kahlhoefer19, Kaplinghat19} that mass loss as a result of tidal stripping may act to accelerate the evolution towards a gravothermal core collapse and eventually black hole formation for dwarf haloes with a self-scattering cross-section of $\sigma_\rmn{T}/m_\rmn{x}$ $\gtrsim$ 5~cm$^{2}$\,g$^{-1}$. Using idealised (i.e. non-cosmological) simulations, \citet{Sameie20} argue that the latter effect might be sufficient to create a diverse subhalo population in a constant cross section SIDM model, e.g., $\sigma_\rmn{T}/m_\rmn{x}$ $=$ 3~cm$^{2}$\,g$^{-1}$, with cored and cuspy subhaloes, depending on whether or not their orbit has triggered the gravothermal collapse.

With the potential for gravothermal collapse in mind, \citet{Zavala2019} simulated a vdSIDM model with a cross-section at relative velocities $<20$~$\rmn{kms}^{-1}$ of 100~cm$^{2}$\,g$^{-1}$ in the Aquarius A halo; this is the highest cross-section value performed in any SIDM $N$-body simulation to date. This comparatively large value of the self-scattering cross-section increases the rate of particle collisions within subhaloes and therefore shortens the time required for gravothermal collapse to within a Hubble time. They showed that this model provides a viable explanation for the broad distribution of cusped and cored density profiles of dark matter subhaloes, and also for the discrepancy between $N$-body simulation predictions on the one hand and observational data on the other: under this model, the classical dwarf spheroidal galaxies are hosted in cored subhaloes whereas the subhaloes in which ultrafaint galaxies reside have instead undergone gravothermal collapse to become cuspy. A similar result has been obtained using a semi-analytic model of gravothermal collapse \citep{Correa20}.

\citep{Zavala2019} briefly presented the existence of a population of very dense haloes in their vdSIDM simulation; in this paper, we study this simulation run in detail, considering the structure of the satellite subhaloes and looking for evidence as to whether the dense subhaloes presented in \citet{Zavala2019} did indeed originate from gravothermal collapse. The structure of this paper is organised as follows. We discuss our simulations in Section~\ref{sec:sim}, present our results in Section~\ref{sec:res} and conclusions in Section~\ref{sec:con}. 

\section{Simulations}
\label{sec:sim}

\begin{figure}
    \centering
    \includegraphics[scale=0.415]{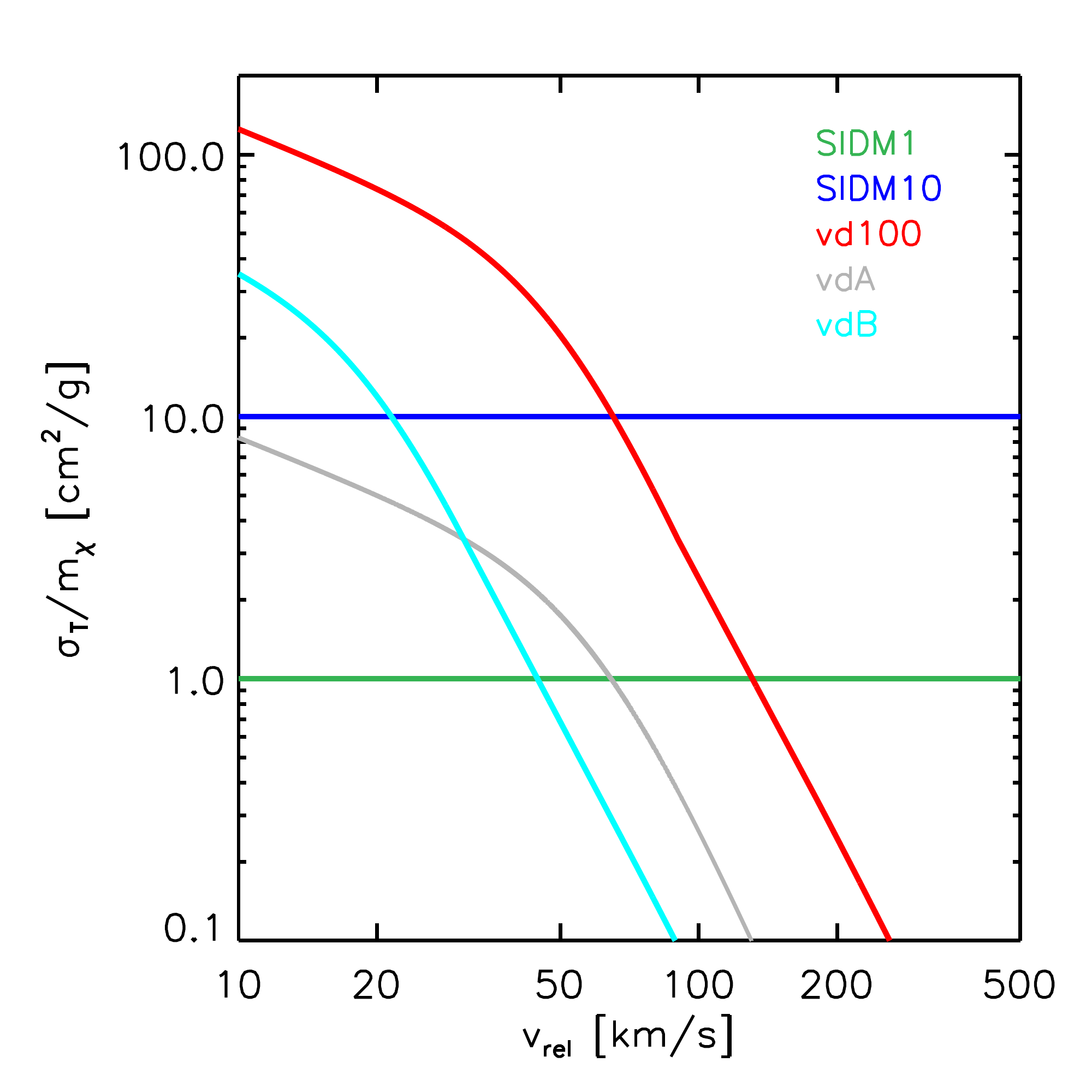}
    \caption{Cross-sections of the SIDM models used in this paper, presented as a function of relative velocity. These models are: SIDM1 (green), SIDM10 (blue), vd100 (red), vdA (grey) and vdB (cyan). }
    \label{fig:cross-section}
\end{figure}

We consider a sample of MW-analogue mass DM-only (DMO) halo simulations, all corresponding to the Aq-A halo first presented in the Aquarius Project \citep{Springel08b}. Most of the simulations were run with the {\sc P-GADGET-3} galaxy formation code \citep{Springel08b}, the one exception is the run introduced by \citet{Zavala2019}. All of the simulations were performed with cosmological parameters consistent with the cosmological parameters of {\it WMAP-1} \citep{wmap1}. The dark matter particle mass and Plummer equivalent gravitational softening lengths are $4.9\times10^{4}$ and M\textsubscript{\(\odot\)} and 120.5~pc respectively; they are resolution level 3 under the Aquarius scheme. Haloes are identified within the friends-of-friends (FoF) algorithm and are decomposed into gravitationally self-bound subhaloes with the {\sc SUBFIND} halo finder code \citep[]{Springel01}.  We require 20 or more bound particles to identify a subhalo. In order to effect comparisons with observations, we select only objects within 300~kpc of the centre of the host halo. The virial mass of the host halo, $M_{200}$, is $1.88\times10^{12}\msun$ in the original CDM simulation, where $M_{200}$ is defined as the mass enclosed within the radius of mean density 200 times the critical density for collapse. The values of $M_{200}$ in all of the simulations are presented in Table~\ref{tab:table}: the variation in this quantity between simulations is less than 5~per~cent.


We present a summary of the vdSIDM simulation first performed in \citet{Zavala2019}. This Aquarius A halo simulation used a vdSIDM model with a self-scattering cross-section, $\sigma_\rmn{T}/m_\rmn{x}$, of 100~cm$^{2}$\,g$^{-1}$ at $V_\rmn{rel}=14$~km~s$^{-1}$ and of 2~cm$^{2}$\,g$^{-1}$ at $V_\rmn{rel}=100$~km~s$^{-1}$ (labelled as vd100). We compare this model to the original CDM Aquarius A halo \citep{Springel08b} and to the suite of SIDM models first presented in \citet{Vogelsberger12}:  velocity independent SIDM models with a cross-section of 1~cm$^{2}$\,g$^{-1}$ (SIDM1) and 10~cm$^{2}$\,g$^{-1}$ (SIDM10); and  vdSIDM models labelled vdA and vdB with parameters $\sigma_\rmn{T}/m_\rmn{x}=6.7$~cm$^{2}$\,g$^{-1}$ and 24~cm$^{2}$\,g$^{-1}$ respectively, both at $V_\rmn{rel}=14$~km~s$^{-1}$). The velocity dependent self-interaction cross-sections are presented in Fig.~\ref{fig:cross-section}, and summaries of the parameters of the six models are given in Table~\ref{tab:table}.



The vdSIDM model is based on a simplified particle physics model where the self-scattering interactions are elastic and driven by an attractive Yukawa-like potential. The self-scattering transfer cross section resulting from this potential can be approximated by a formula originally used in plasma physics, and introduced in \citet{Feng10, Loeb11}, in the context of SIDM. The assignment of a post-scattering particle velocity maintains the conservation of both linear momentum and energy (see \citealp{Vogelsberger12} for the algorithm implementation of SIDM used in the simulations in Table~\ref{tab:table}). 

\begin{figure*}
    \centering
    
     \setbox1=\hbox{\includegraphics[scale=0.16]{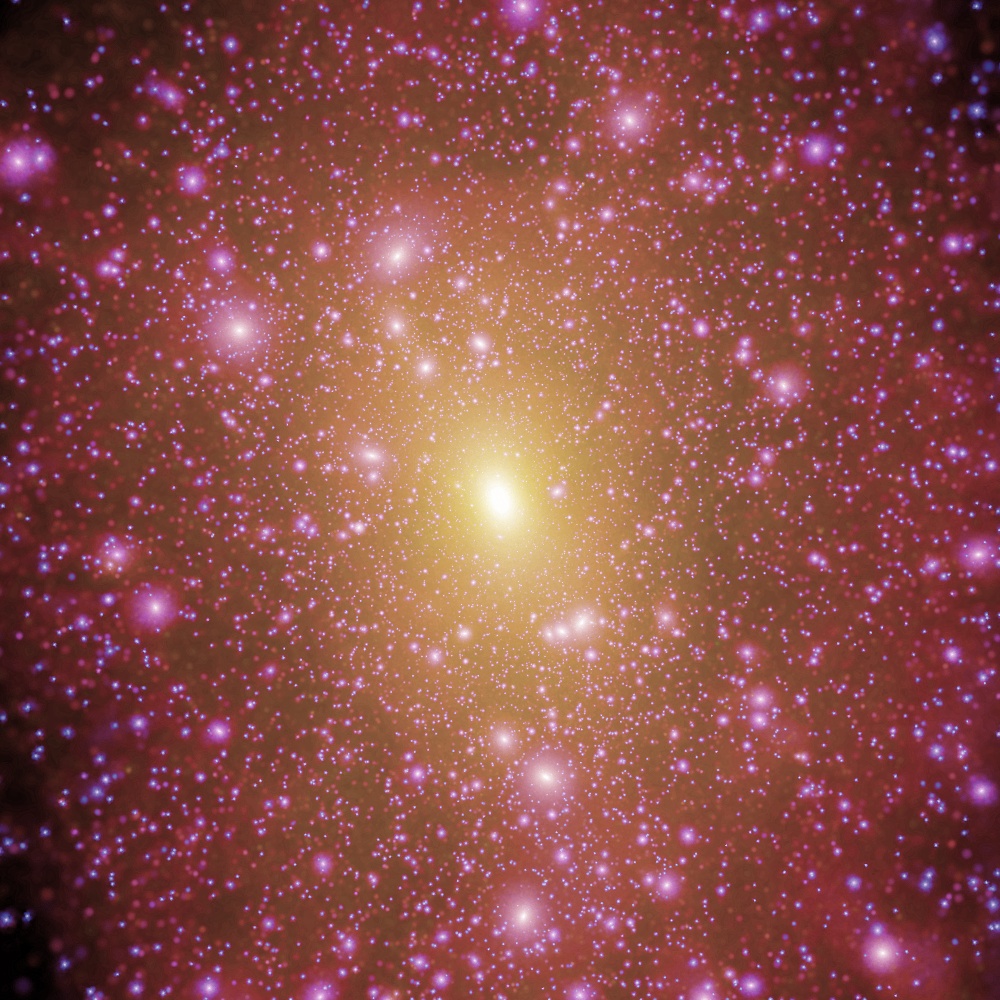}}
    
     \includegraphics[scale=0.16]{AqA3RefP0_BSize0p4Mpc2_X1Y2_001.jpg}\llap{\makebox[\wd1][l]{\raisebox{0.68\wd1}{\includegraphics[width=0.28\textwidth]{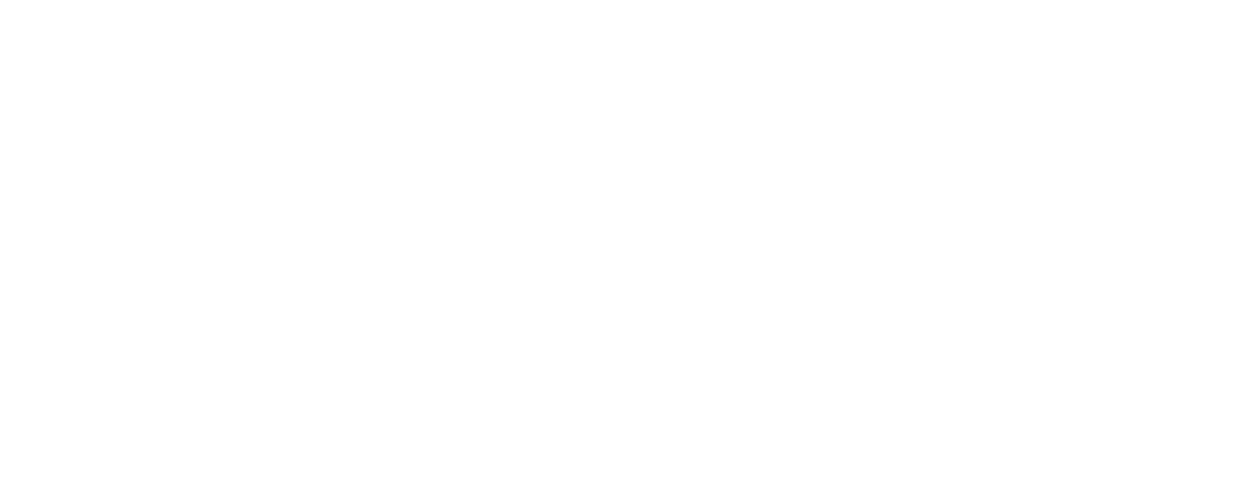}}}}
      \includegraphics[scale=0.16]{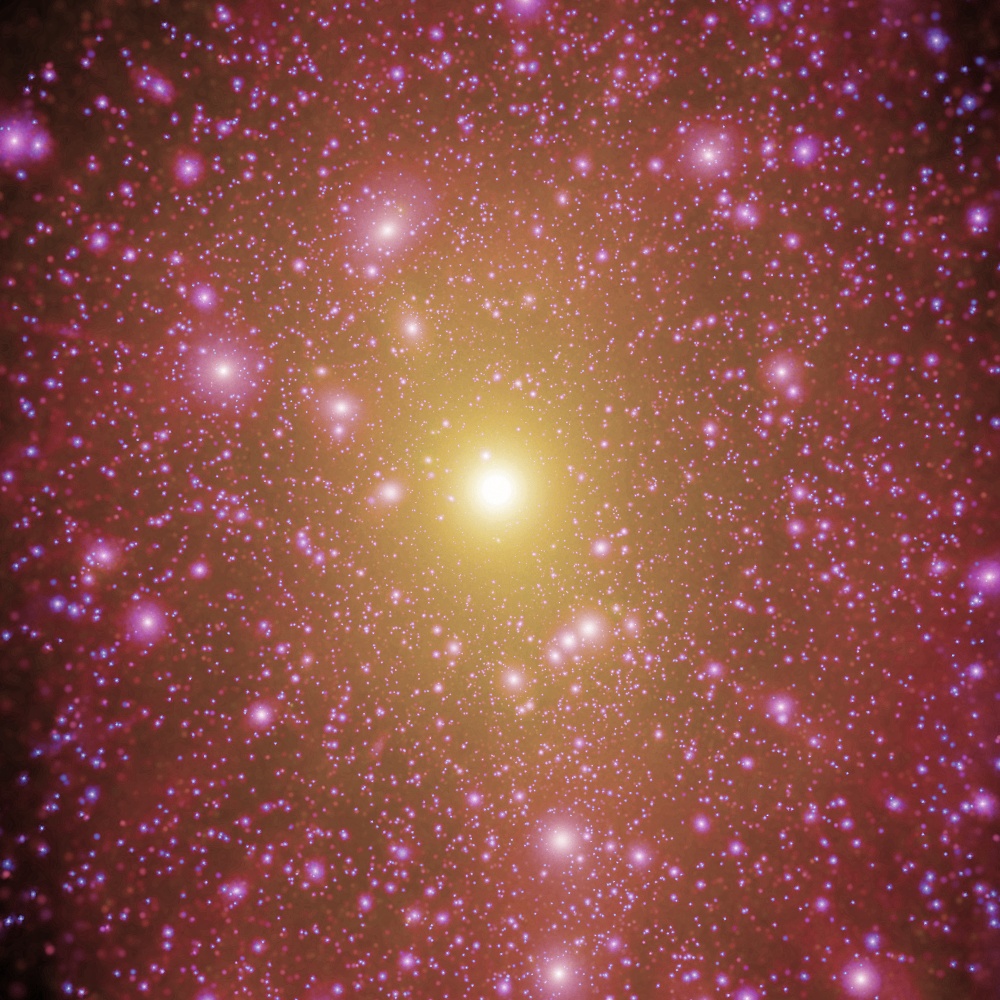}\llap{\makebox[\wd1][l]{\raisebox{0.68\wd1}{\includegraphics[width=0.28\textwidth]{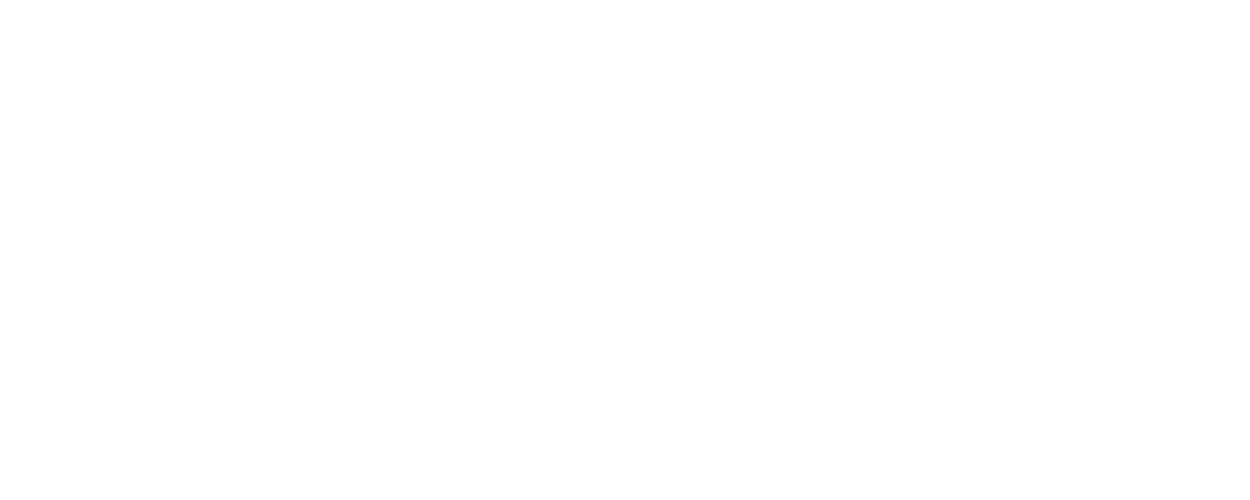}}}} 
       \includegraphics[scale=0.16]{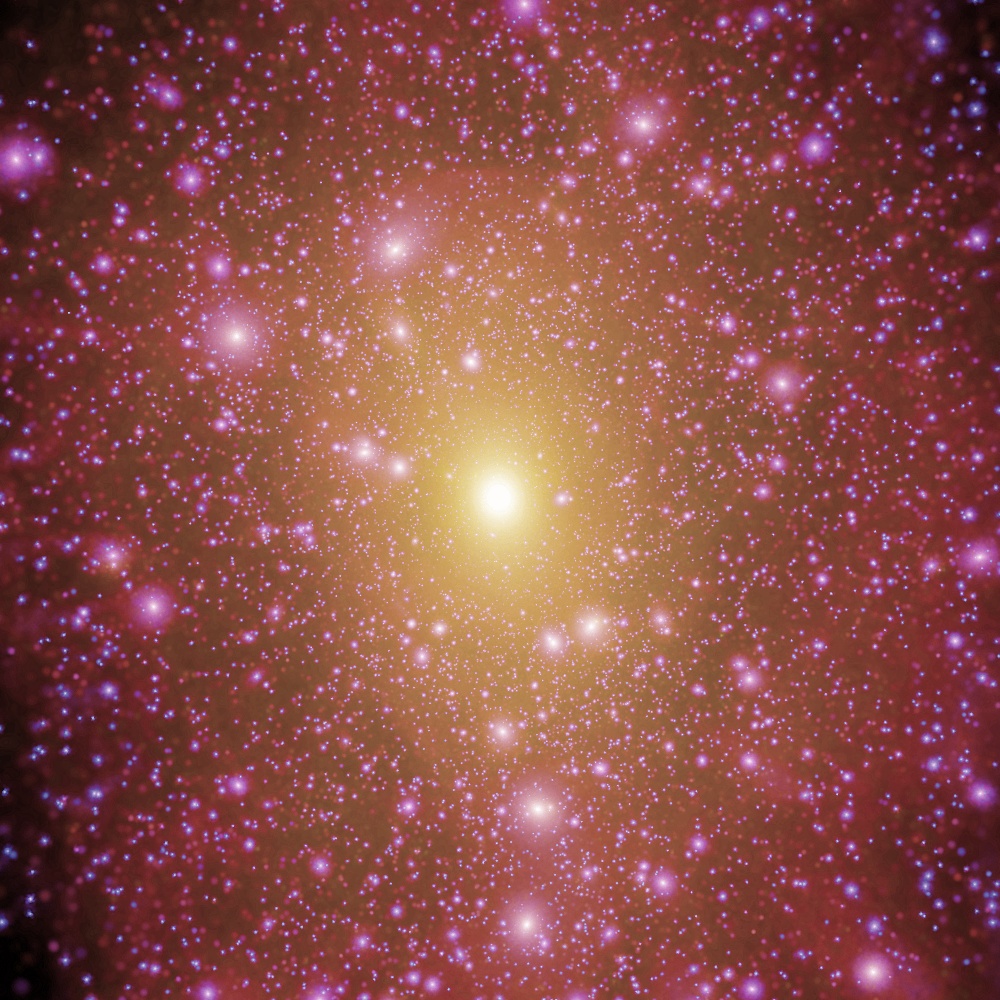}\llap{\makebox[\wd1][l]{\raisebox{0.68\wd1}{\includegraphics[width=0.28\textwidth]{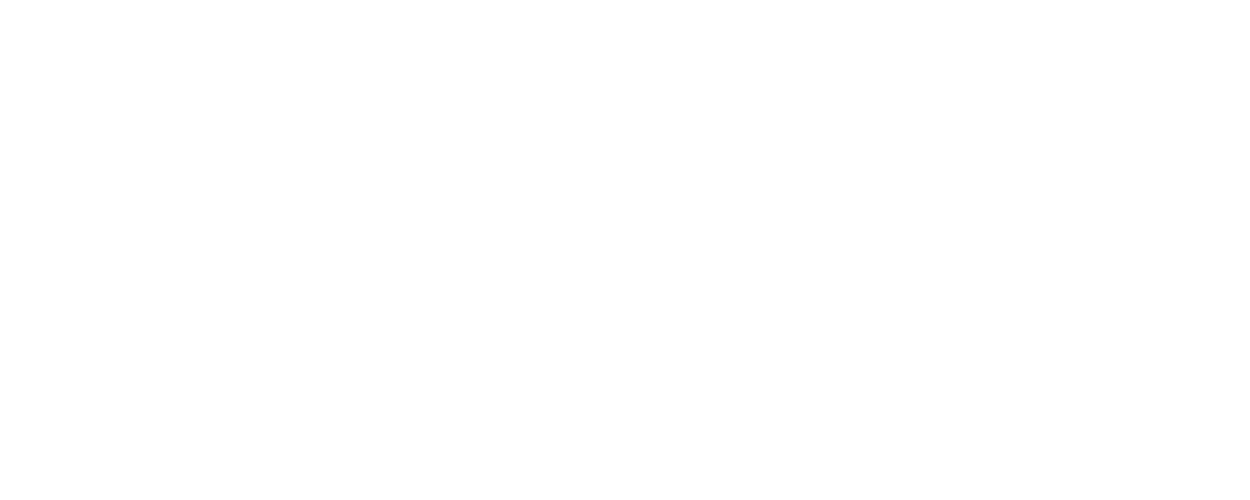}}}}\\
        \vspace{-1mm}
      \includegraphics[scale=0.16]{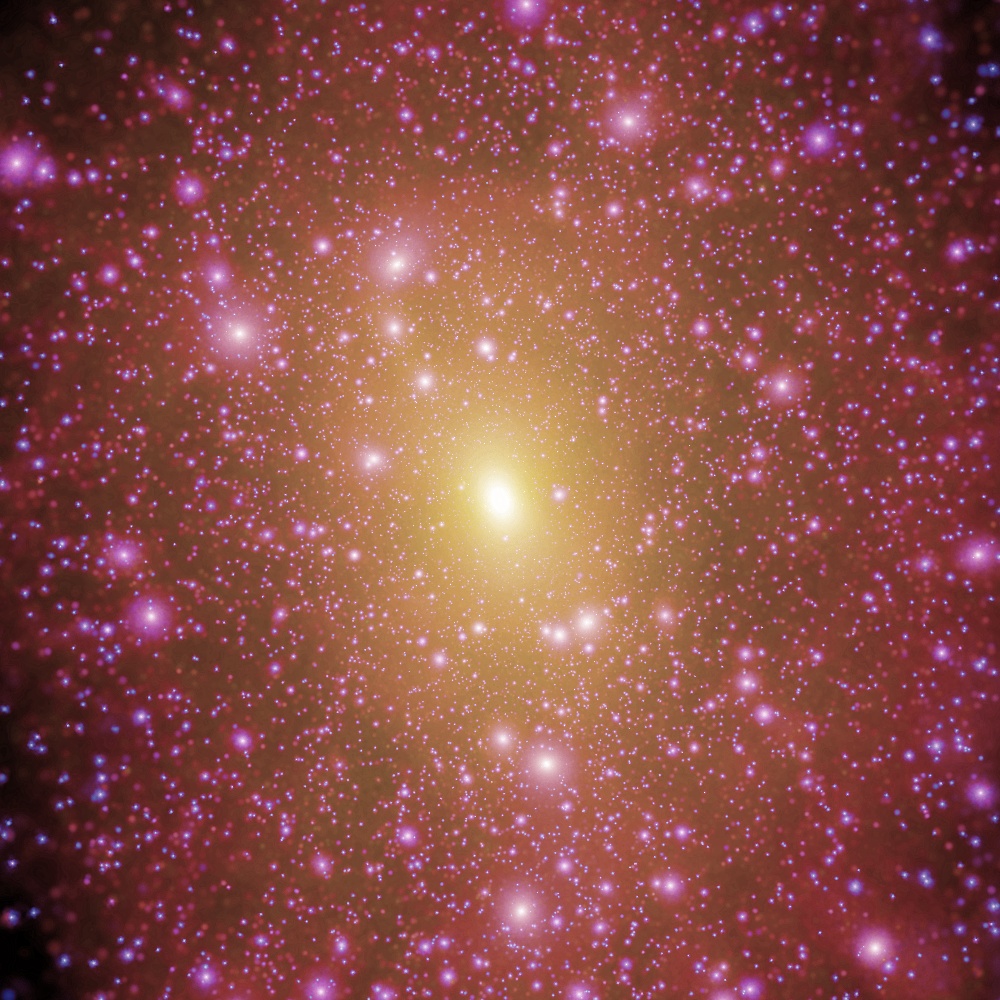}\llap{\makebox[\wd1][l]{\raisebox{0.68\wd1}{\includegraphics[width=0.28\textwidth]{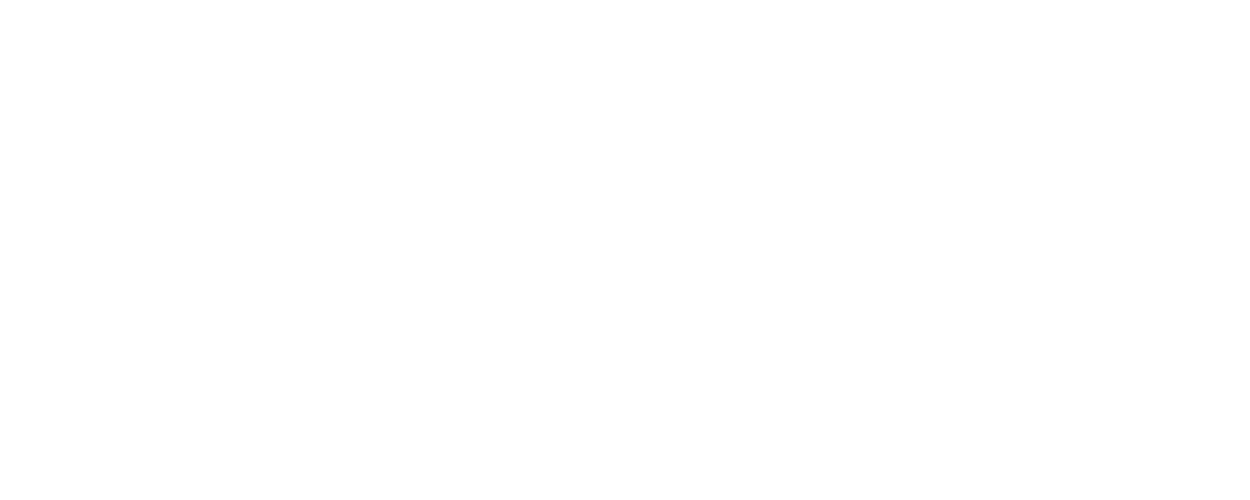}}}}
      \includegraphics[scale=0.16]{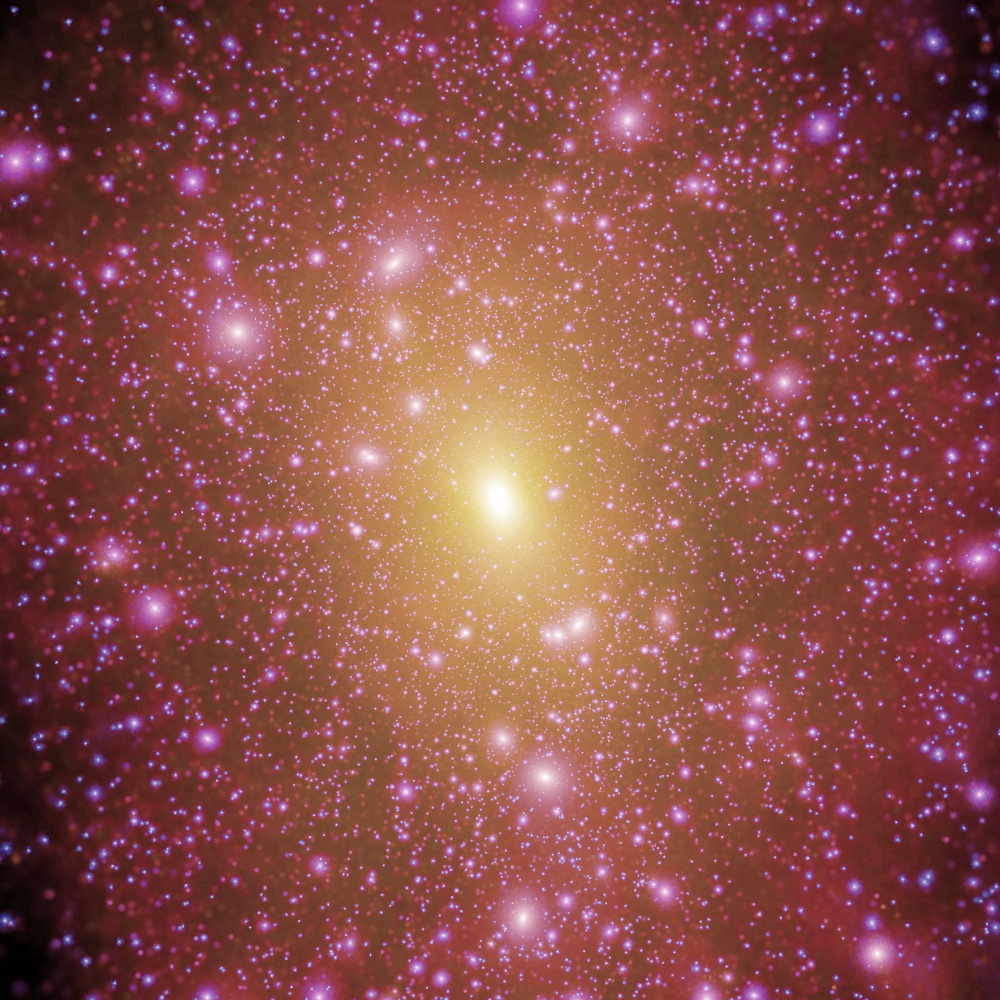}\llap{\makebox[\wd1][l]{\raisebox{0.68\wd1}{\includegraphics[width=0.28\textwidth]{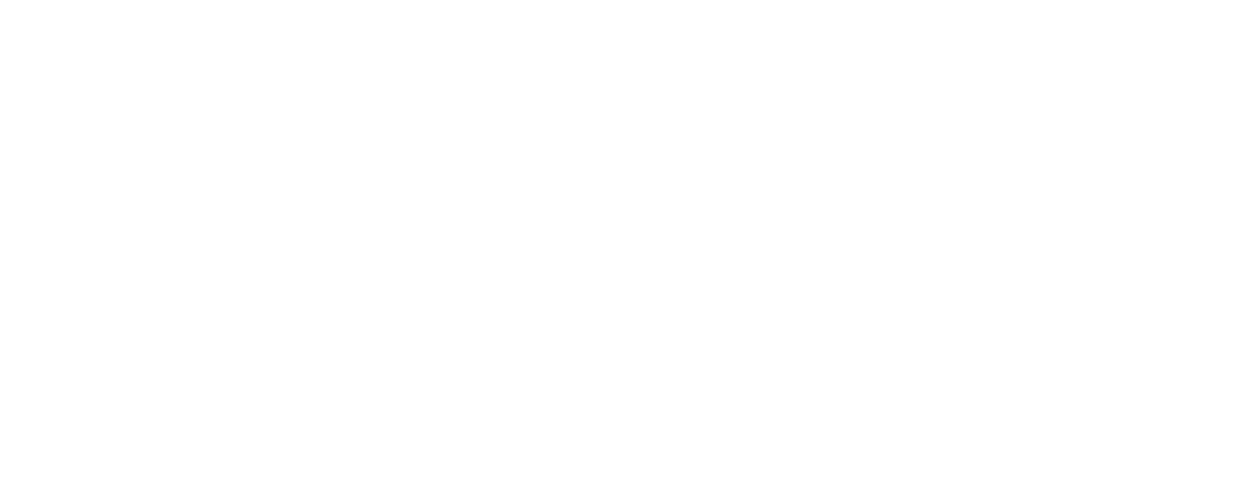}}}} 
      \includegraphics[scale=0.16]{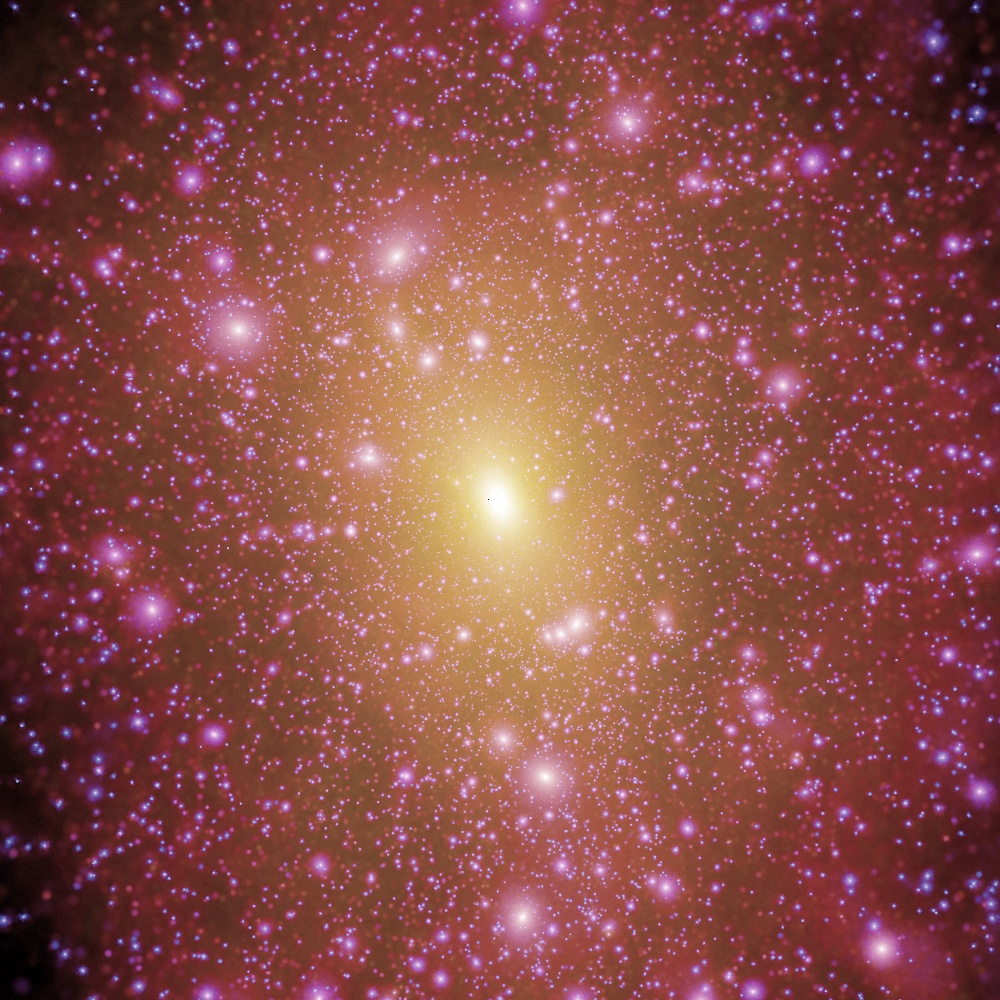}\llap{\makebox[\wd1][l]{\raisebox{0.68\wd1}{\includegraphics[width=0.28\textwidth]{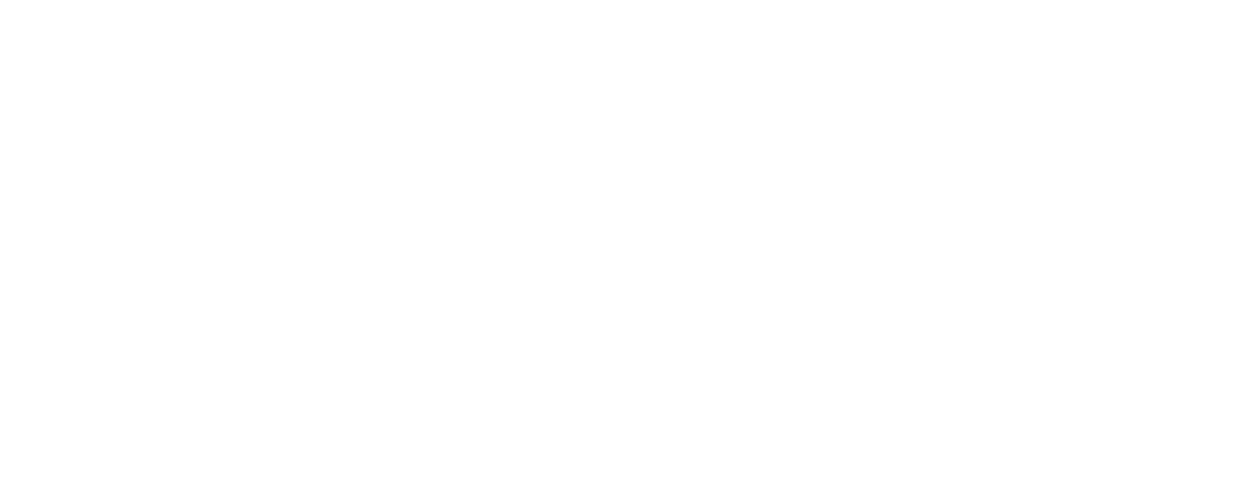}}}}\\
    
    \caption{Images of the six haloes used in this study. The image intensity correlates with the dark matter column density and the the colour traces the velocity dispersion. Each image is 400~kpc on a side. The six simulations, from left to right and top to bottom, are: CDM, SIDM10, SIDM1, vd100, vdB and vdA.}
    \label{fig:HaloImagesHN}
\end{figure*}

We present images of the simulations used in this study to provide an illustration of how the effect caused by the change in dark matter model. Fig.~\ref{fig:HaloImagesHN} shows images of the six haloes, where the image intensity indicates the dark matter column density and the image hue is related to the velocity dispersion. The most striking difference is how the SIDM10 model results in a halo that is markedly more spherical than is the case in CDM. It is also apparent that the abundance of subhaloes in the centre of the SIDM10 suppression is suppressed, which is due to evaporation as shown by \citep{Vogelsberger12}. SIDM1 constitutes an intermediate case between SIDM10 and CDM. By contrast, all three vdSIDM models have very similar appearances to CDM down to the precise locations of the more massive subhaloes. 

\begin{figure*}
    \centering
    
     \setbox1=\hbox{\includegraphics[scale=0.16]{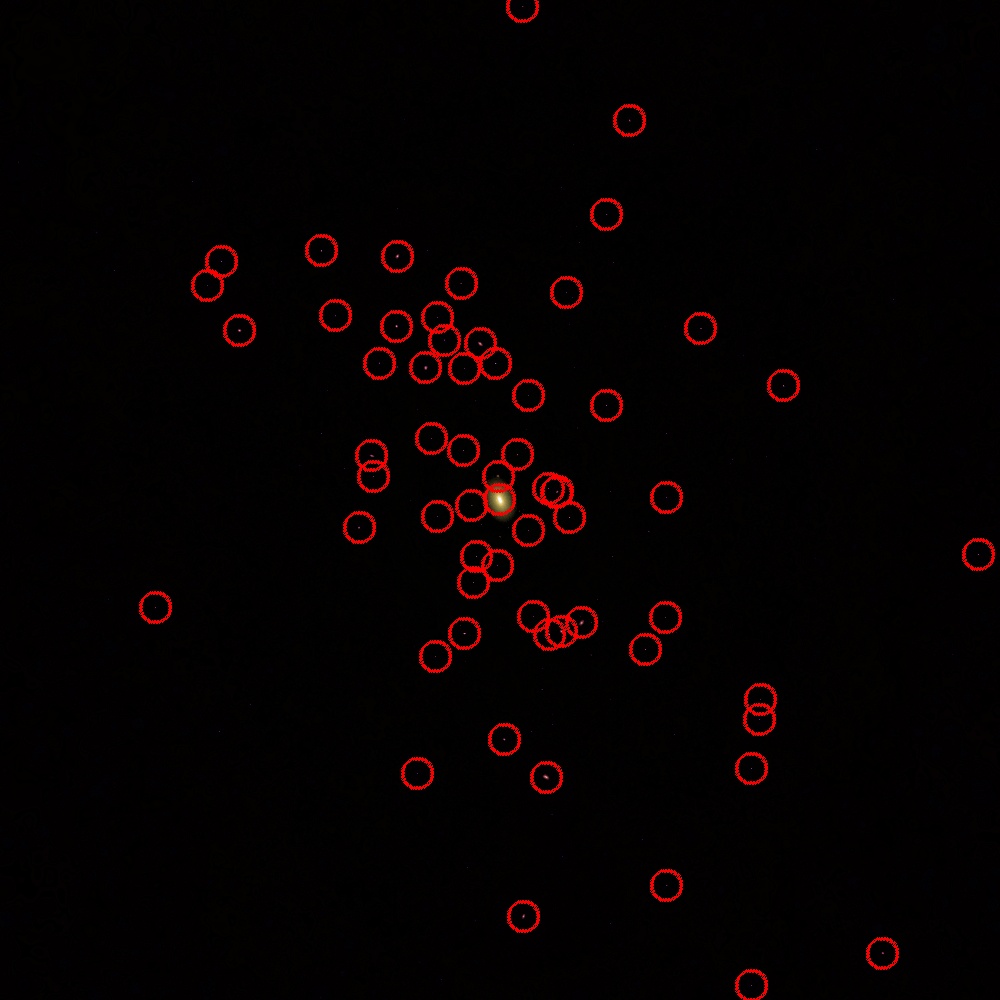}}
    
     \includegraphics[scale=0.16]{AqA3RefP0_BSize0p4Mpc2_X1Y2_HighDens_001.jpg}\llap{\makebox[\wd1][l]{\raisebox{0.68\wd1}{\includegraphics[width=0.28\textwidth]{A3CDM_Label-eps-converted-to.pdf}}}}\llap{\makebox[\wd1][l]{\raisebox{-0.1\wd1}{\includegraphics[width=0.25\textwidth]{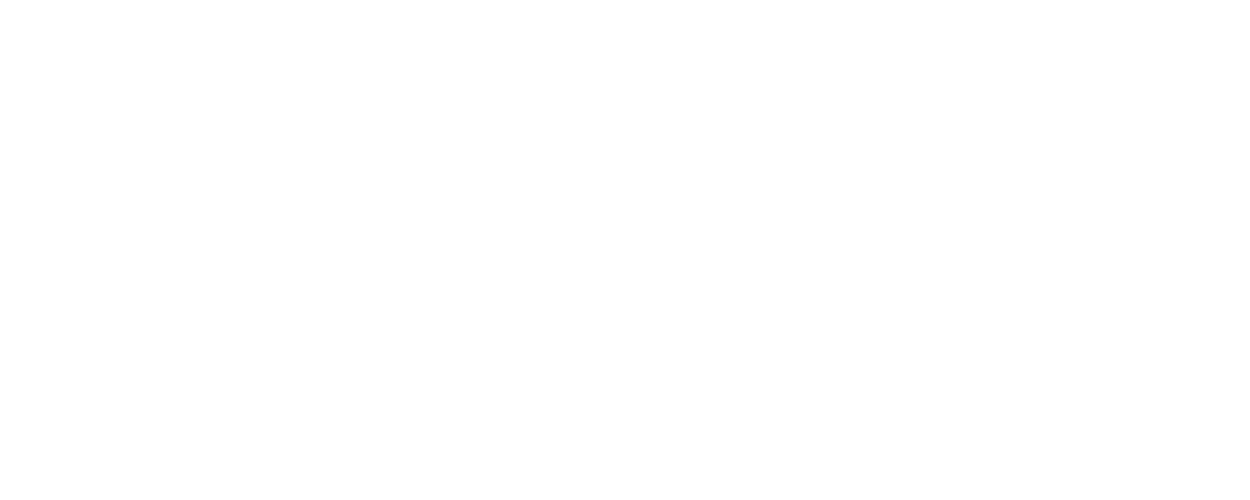}}}}
      \includegraphics[scale=0.16]{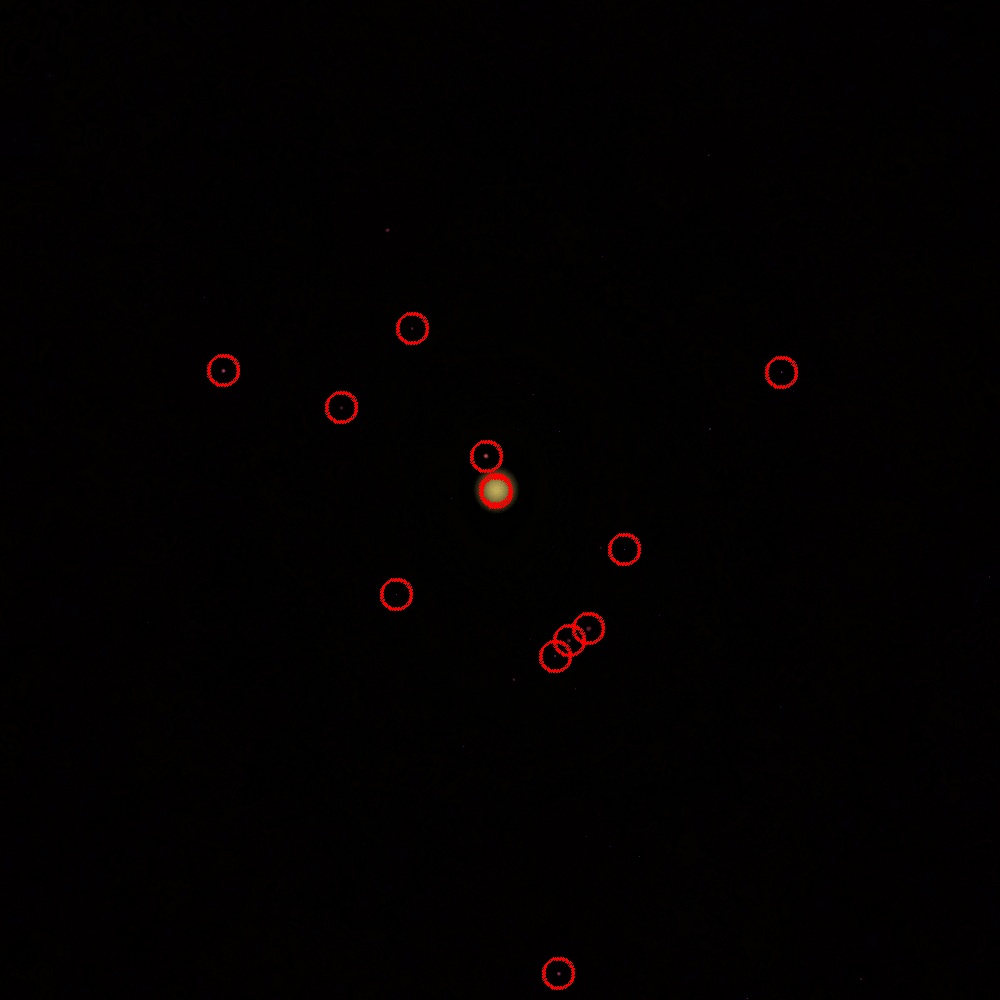}\llap{\makebox[\wd1][l]{\raisebox{0.68\wd1}{\includegraphics[width=0.28\textwidth]{A3RefP1_Label-eps-converted-to.pdf}}}}\llap{\makebox[\wd1][l]{\raisebox{-0.1\wd1}{\includegraphics[width=0.25\textwidth]{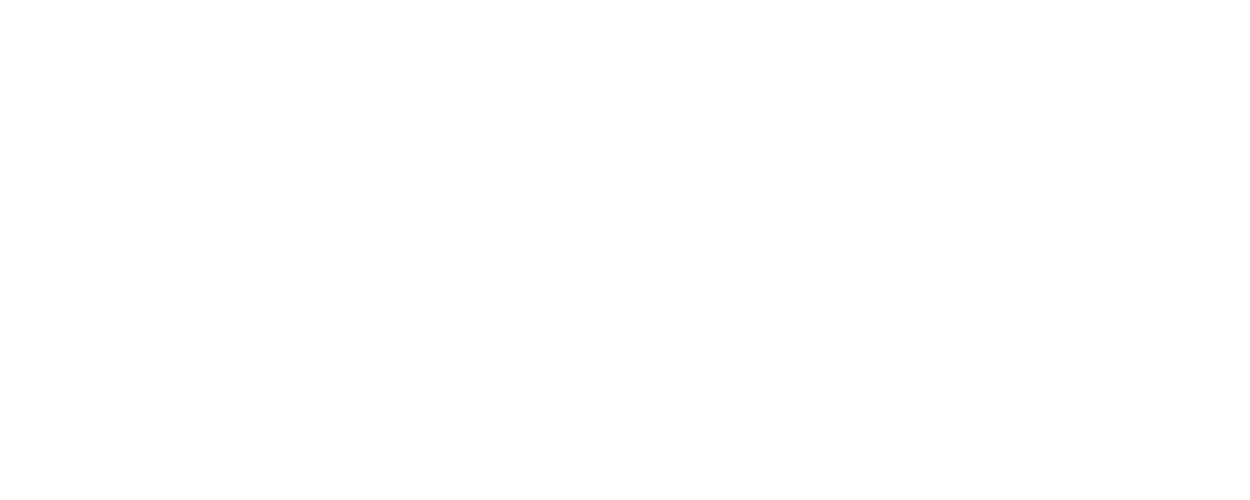}}}} 
       \includegraphics[scale=0.16]{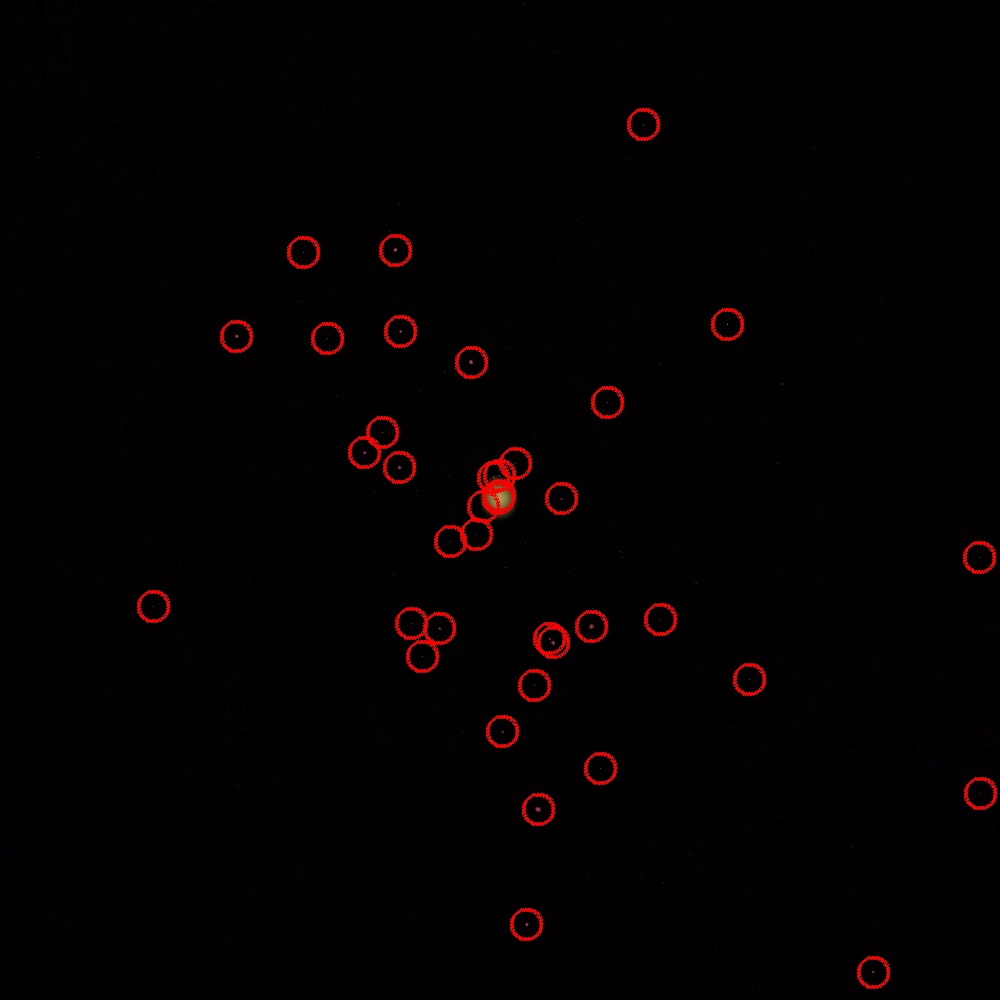}\llap{\makebox[\wd1][l]{\raisebox{0.68\wd1}{\includegraphics[width=0.28\textwidth]{A3RefP4_Label-eps-converted-to.pdf}}}}\llap{\makebox[\wd1][l]{\raisebox{-0.1\wd1}{\includegraphics[width=0.25\textwidth]{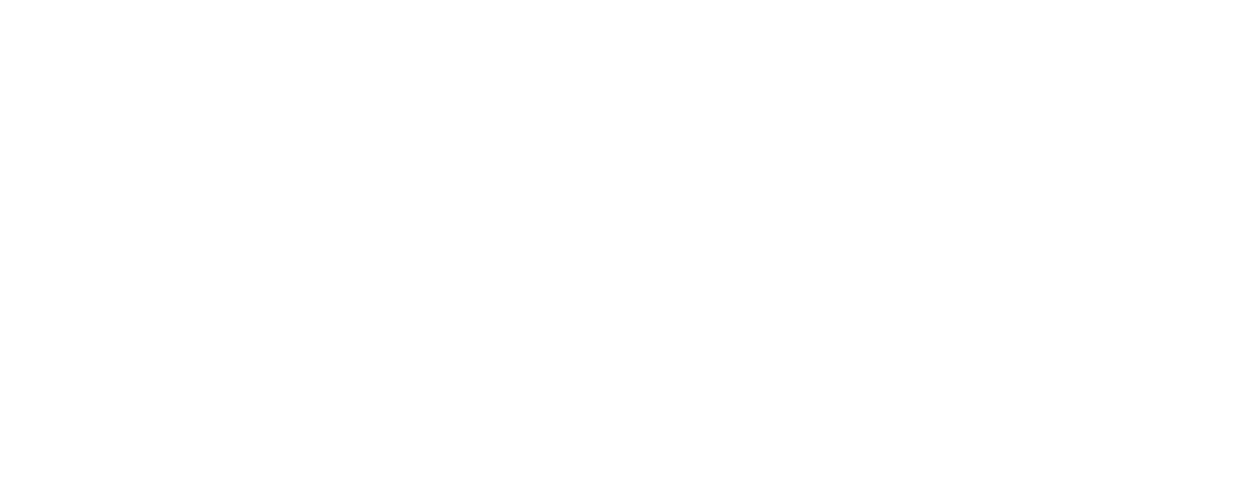}}}}\\
        \vspace{-7mm}
      \includegraphics[scale=0.16]{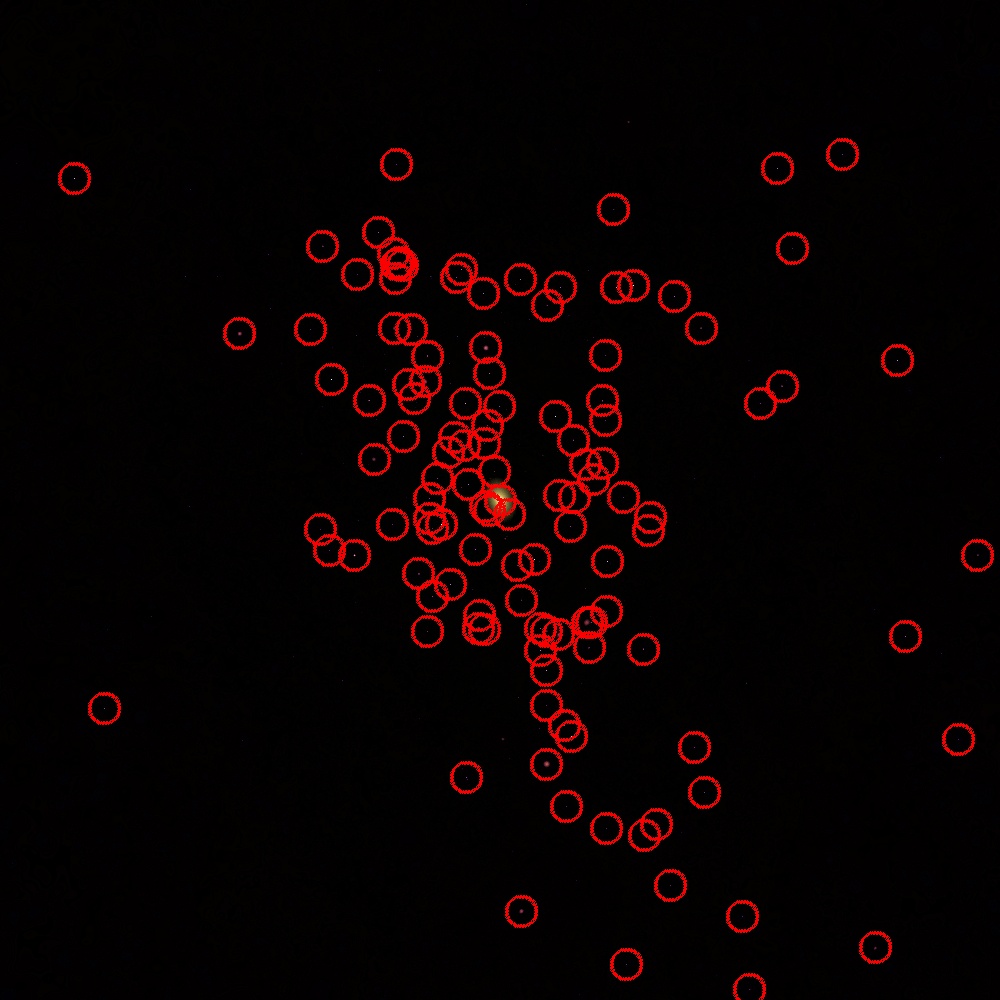}\llap{\makebox[\wd1][l]{\raisebox{0.68\wd1}{\includegraphics[width=0.28\textwidth]{A3vd100_Label-eps-converted-to.pdf}}}}\llap{\makebox[\wd1][l]{\raisebox{-0.1\wd1}{\includegraphics[width=0.25\textwidth]{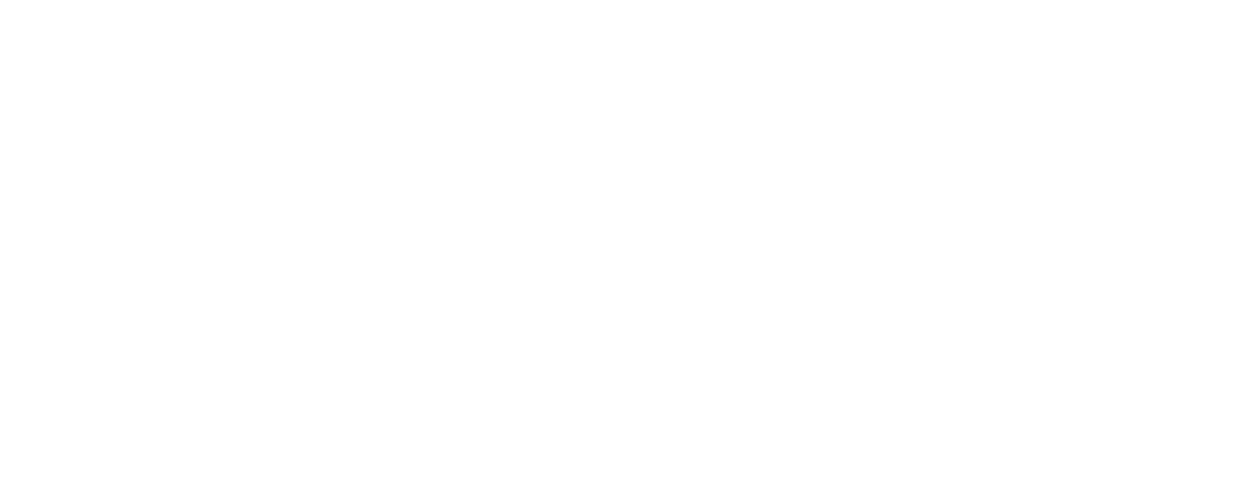}}}}
      \includegraphics[scale=0.16]{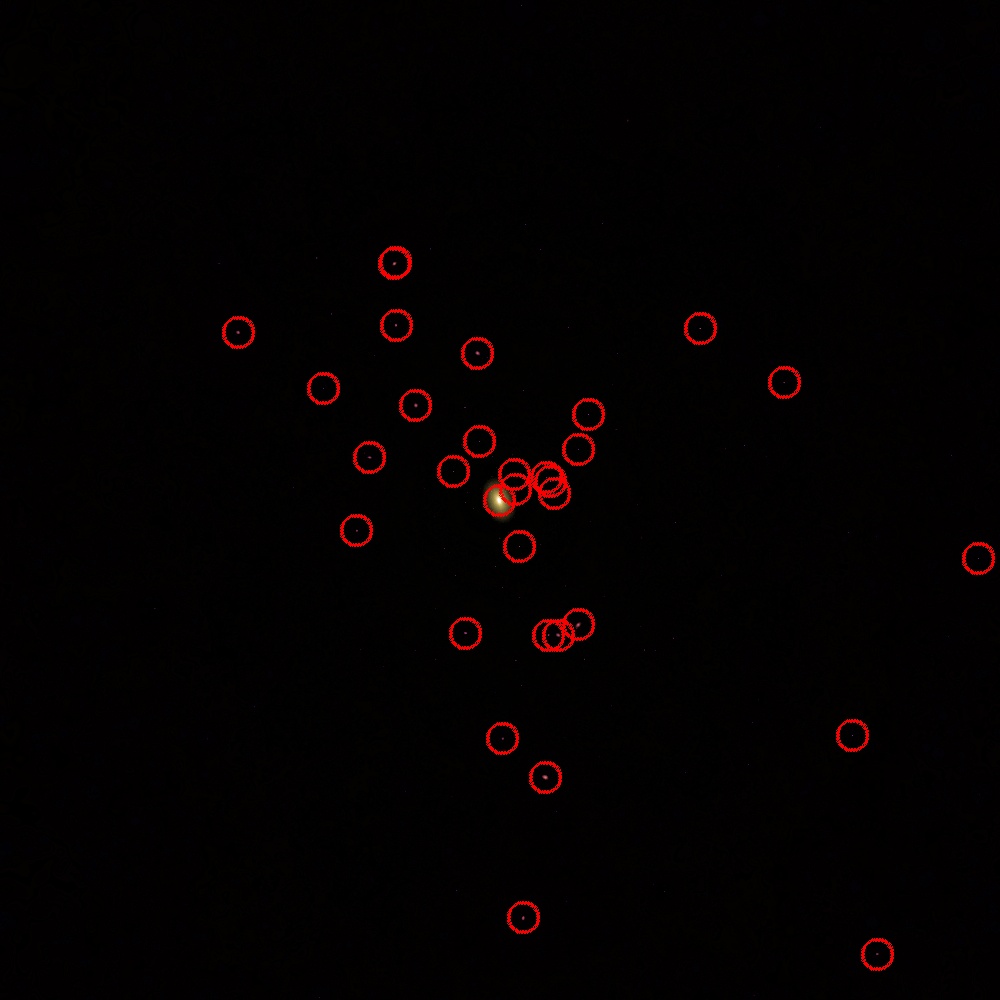}\llap{\makebox[\wd1][l]{\raisebox{0.68\wd1}{\includegraphics[width=0.28\textwidth]{A3RefP3_Label-eps-converted-to.pdf}}}}\llap{\makebox[\wd1][l]{\raisebox{-0.1\wd1}{\includegraphics[width=0.25\textwidth]{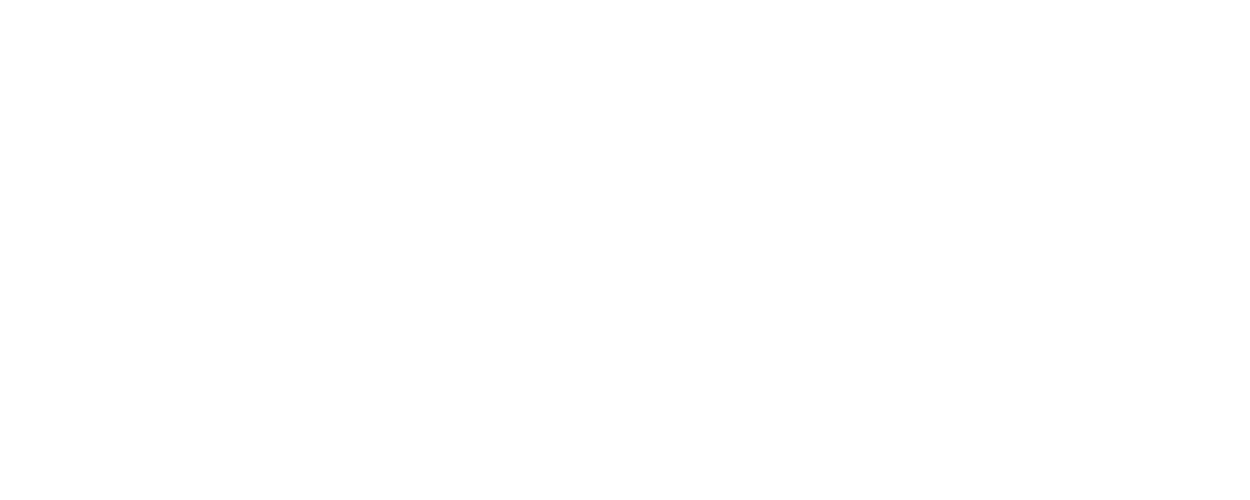}}}} 
      \includegraphics[scale=0.16]{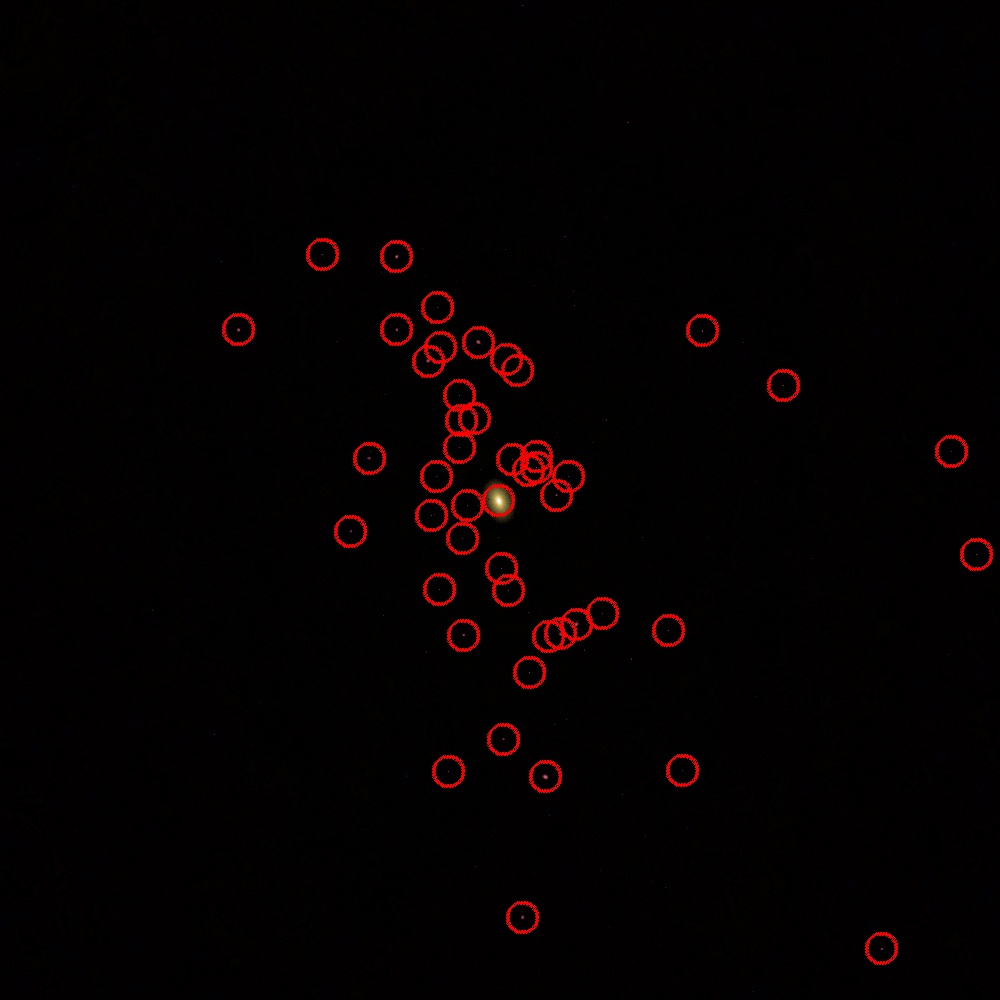}\llap{\makebox[\wd1][l]{\raisebox{0.68\wd1}{\includegraphics[width=0.28\textwidth]{A3RefP2_Label-eps-converted-to.pdf}}}}\llap{\makebox[\wd1][l]{\raisebox{-0.1\wd1}{\includegraphics[width=0.25\textwidth]{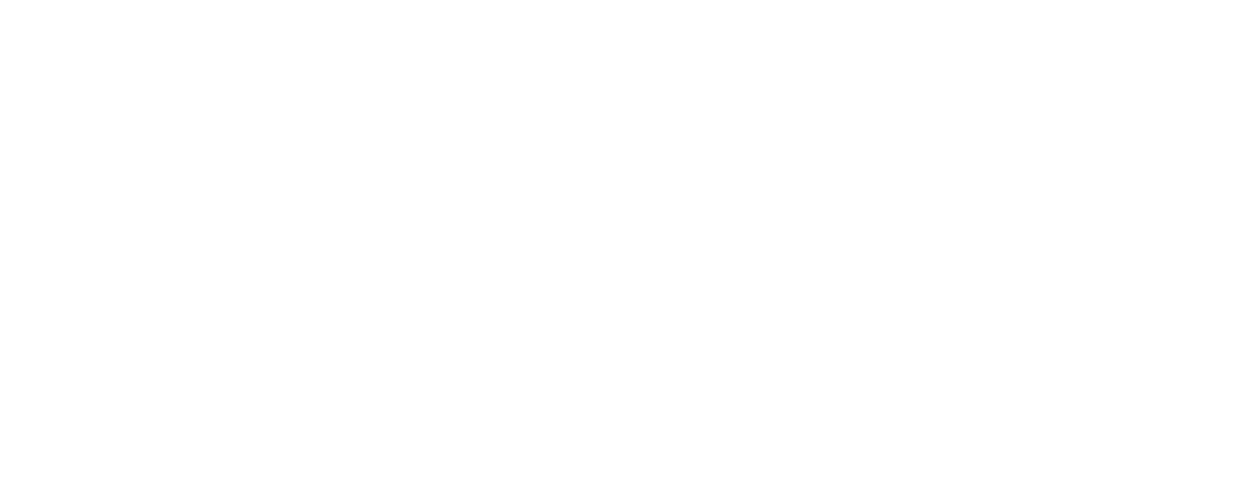}}}}\\
    
    \caption{ Images of the highest density peaks in the six haloes used in this study. 
    The density range used to make each image is restricted to pixels with a higher integrated column density than 99.92~per~cent of CDM pixels; the velocity dispersion-colour relation is the same as in Fig.~\ref{fig:HaloImagesHN}. Each density peak that passes the filter is enclosed with a red circle, and the number of such peaks is given in the panel. Each image is 400~kpc on a side. The six simulations, from left to right and top to bottom, are: CDM, SIDM10, SIDM1, vd100, vdB and vdA.}
    \label{fig:HaloImages}
\end{figure*}

Given that most SIDM models are tailored to alter the density profiles of haloes without having a measurable impact on the halo mass function, we would anticipate that only regions of very high density exhibit differences between CDM and the vdSIDM simulations, either in their abundance or in their spatial distribution. Therefore, we recompute the images of Fig.~\ref{fig:HaloImagesHN} with a very high threshold density filter, in order to present the number and distribution of high density peaks (in projection). We present these modified images in Fig.~\ref{fig:HaloImages}; the red circles mark sets of pixels in which the integrated column density is higher than that of 99.92~per~cent of pixels in the CDM image, and we refer to these regions as `dense peaks'. 

The CDM and vdSIDM models result in host haloes with a central cusp, whilst the velocity independent SIDM models result in a halo core as conveyed by the lower image intensity in the latter two simulation images. The number of dense peaks in SIDM10 is only a third or lower of any of the other simulations for two reasons: first, the higher cross-section  {\it independently} of the velocity of the particles evaporates some subhaloes (due to collisions between particles in the subhalo with those in the host), and second, the self-interactions between particles in the centre of subhaloes carve out lower density cores in the subhaloes that do not evaporate. In contrast to Fig.~\ref{fig:HaloImagesHN}, there are clear differences between the vdSIDM models and CDM. vdB shows only half as many peaks as CDM, as a result of core creation by its large maximum cross-section, and vdA has only 50~per~cent more peaks then vdB. Most notably, the vd100 run shows twice as many as even CDM; a first piece of evidence that a fraction of vd100 subhaloes are undergoing gravothermal collapse, with cores collapsing to form very high density, cuspy profiles.

\begin{table*}
	\centering
	\caption{Properties of the models considered in this paper: simulation name; the cosmology of the simulation; the maximum self-scattering transfer cross-section per unit mass (cm$^{2}$\,g$^{-1}$); the simulation dark matter particle mass ($m_\rmn{DM}$); host halo mass ($M_\rmn{halo}$); the Plummer equivalent gravitational softening length ($\epsilon$); and the original reference describing each simulation. For the velocity dependent models we also supply the relative velocity $(V\textsubscript{rel})$ under the cross-section heading.}
	\begin{tabular}{lcccccl}
		\hline
		Name & Cosmology & $\sigma_\rmn{T}/m_\rmn{x}$ (cm$^{2}$\,g$^{-1}$) & $m_\rmn{DM}$ (M\textsubscript{\(\odot\)}) & $M_\rmn{200}$ ($10^{12}$ M\textsubscript{\(\odot\)}) & $\epsilon$ (pc) & Reference\\
		\hline
		CDM & WMAP-1 & - & $4.9\times10^{4}$ & 1.88 & 120.5 & \citet{Springel08b}\\
		SIDM1 & WMAP-1 & 1 & $4.9\times10^{4}$ & 1.90 & 120.5 & \citet{Zavala13}\\
		SIDM10 & WMAP-1 & 10 & $4.9\times10^{4}$ & 1.82 & 120.5 &
		\citet{Vogelsberger12}\\
		vdA & WMAP-1 & 6.7 (V\textsubscript{rel}~=~14~km~s$^{-1}$) & $4.9\times10^{4}$ & 1.91 & 120.5 & \citet{Vogelsberger12}\\
		vdB & WMAP-1 & 24 (V\textsubscript{rel}~=~14~km~s$^{-1}$) & $4.9\times10^{4}$ & 1.91 & 120.5 & \citet{Vogelsberger12}\\
		vd100 & WMAP-1 & 100 (V\textsubscript{rel}~=~14~km~s$^{-1}$) & $4.9\times10^{4}$ & 1.84 & 120.5 & \citet{Zavala2019}\\
		\hline
	\end{tabular}

    \label{tab:table}
\end{table*}

\section{Results}
\label{sec:res}

\subsection{Subhaloes at $z=0$}

\subsubsection{The subhalo population}

\begin{figure}
    \includegraphics[scale=0.22,angle=0]{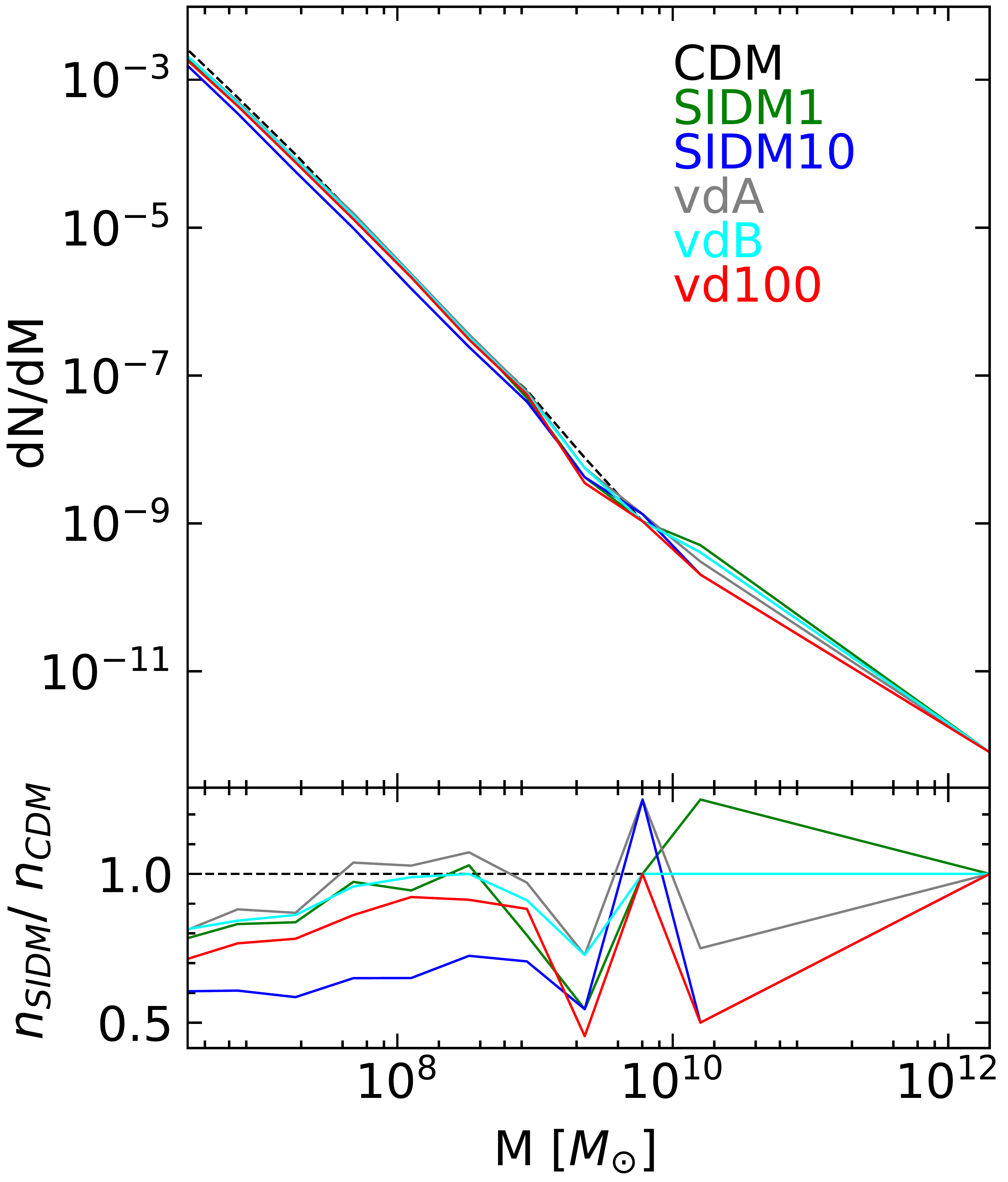}
    \caption{Differential subhalo abundance as a function of enclosed mass for all dark matter models considered. The SIDM1, SIDM10, vd100, vdA and vdB models are denoted as green, blue, red, grey and cyan solid lines respectively. The CDM model is denoted by a black dashed line. The bottom panel shows the mass function of all SIDM models as a ratio to that of the CDM model.}
    \label{fig:mass_1} 
\end{figure}

We begin our presentation of the results with a discussion of the subhalo mass function, for subhaloes located within 300~kpc of the host halo centre irrespective of their FoF group membership. We define the subhalo mass, $M_\rmn{sub}$, as the total mass gravitationally bound to each subhalo as determined by {\sc subfind}.  Fig.~\ref{fig:mass_1} shows the differential subhalo mass functions for each of the six simulations, including a second panel that shows the ratio of the SIDM mass functions relative to CDM. The approximate shape of the mass function is the same for all six models in the mass range $[10^{8},10^{10}]$~$\msun$, which is consistent with the results of previous studies \citep[e.g.][]{Vogelsberger12, Rocha13, Zavala13}. The bottom panel of Fig.~\ref{fig:mass_1} shows the ratio of SIDM subhalo counts in relation to that of the CDM model while factoring out the steepness of the mass function. All of the SIDM models predict fewer low mass subhaloes ($<4\times10^{7}$~$\msun$) than the CDM model. The suppression is strongest for SIDM10, at 60~per~cent of the CDM halo mass function due to the evaporation demonstrated by \citet{Vogelsberger12}. Of the other SIDM models, vd100 model predicts the fewest subhaloes, with 90~per~cent of CDM at $M_\rmn{sub}\sim10^{8}$~$\msun$ and 75~per~cent at $M_\rmn{sub}\sim10^{7}$~$\msun$. This demonstrates part of the way in which the vdSIDM model addresses the TBTF problem by predicting fewer subhaloes capable of hosting luminous galaxies, with the aforementioned mass range being synonymous with the dark matter halo mass of `problematic' dwarf galaxies such as Tucana \citep{Gregory19}. 

\begin{figure}
    \includegraphics[scale=0.19,angle=0]{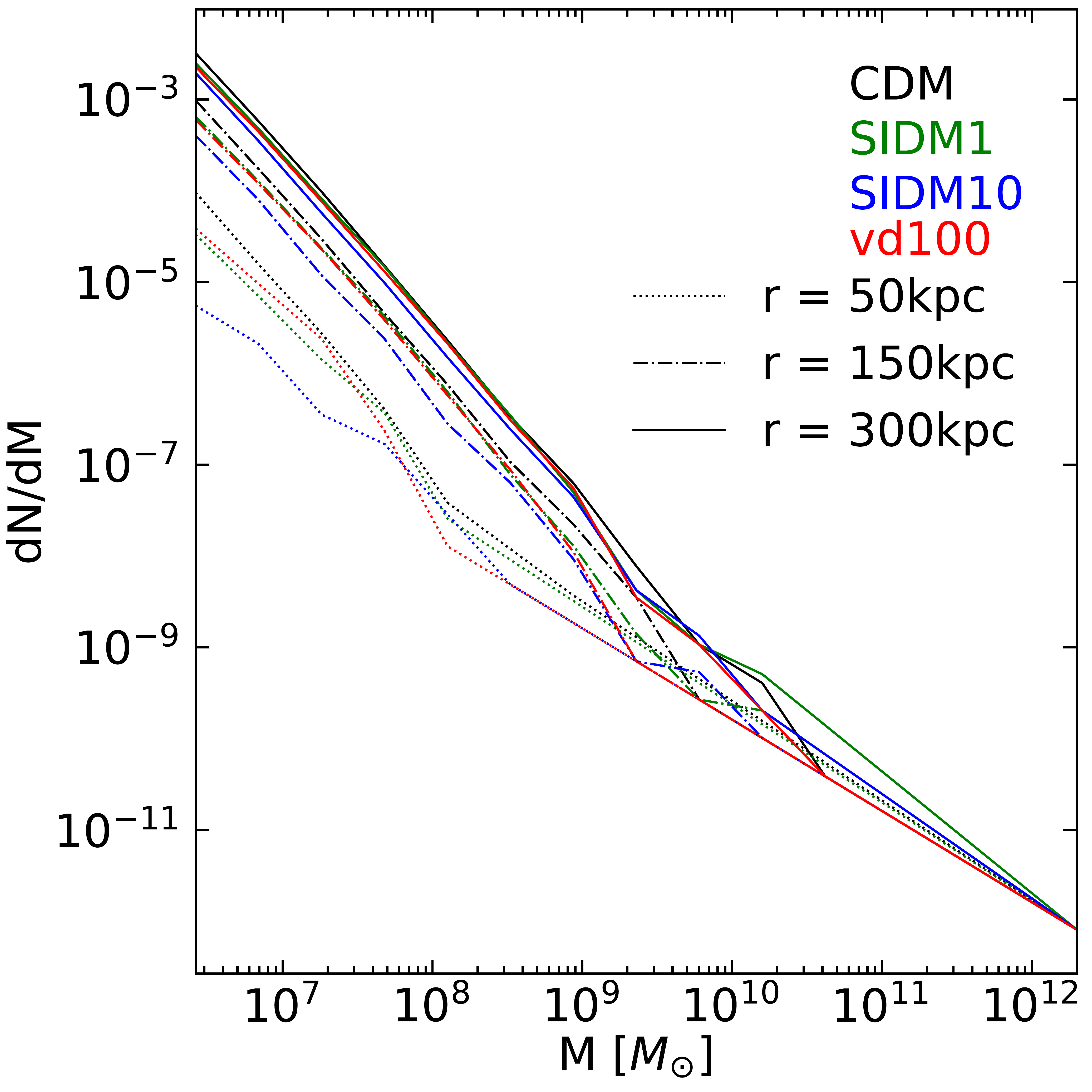}  
    \caption{Differential subhalo mass function for all dark matter models considered, binned by radii from the centre of the host halo. Our three radial bins, <~50~kpc, <~150~kpc and <~300~kpc are denoted by dotted, dot-dashed and solid lines, respectively. The SIDM1, SIDM10 and vd100 models are denoted as green, blue and red solid lines respectively. The CDM model is denoted by black lines.}
    \label{fig:mass_rad} 
\end{figure}

Of particular interest in the properties of subhaloes in the vdSIDM models is the change in the subhalo mass function between different distances from the host halo centre. Previous studies have shown that subhalo evaporation is most pronounced close to the host halo centre \citep{Vogelsberger12} and others have argued that close orbit pericentres hasten the onset of gravitational collapse \citep{Nishikawa20}. A careful study would require that we follow the orbits of subhaloes with more time resolution than we have available in our stored output. We therefore instead compute halo mass functions for three radial bins: in addition to all haloes within 300~kpc as featured in Fig.~\ref{fig:mass_1}, we consider all subhaloes within 150~kpc of the host centre and all haloes within 50~kpc, and present the results in  Fig.~\ref{fig:mass_rad}. For clarity we omit the vdA and vdB simulations from this figure, and thus only include CDM, SIDM1, SIDM10, and vd100.

The difference between the four models is strongest in the smallest radial bin, especially between SIDM10 and CDM as was expected for the combination of such a large, velocity independent cross-section on the one hand and the dense inner host halo environment on the other. The vd100 model shows a much smaller difference, which is consistent with previous claims that subhalo destruction from high velocity collisions with the parent halo at a higher self-scattering cross-section is mitigated when a velocity dependence of the cross-section is introduced \citep[]{Zavala2019}. It instead produces approximately the same number of subhaloes as SIDM1.

The value of the bound subhalo mass is sensitive to the halo finder \citep{Onions12}. It has therefore been common to use a further parameter as a proxy for mass: the maximum circular velocity, $V_\rmn{max}$. This measurement refers to the velocity of a test particle moving in a circular orbit around a subhalo, and is given by:

\begin{equation}
    V_\rmn{max}^2=\frac{GM(r_\rmn{max})}{r_\rmn{max}},
	\label{eq:CircVel}
\end{equation}
\noindent
where $G$ is the universal gravitational constant, $M(r)$ is the enclosed bound mass of the subhalo as a function of radius and $r_\rmn{max}$ is the radius from the centre of the subhalo that corresponds to $V_\rmn{max}$. This parameter has historically been a useful proxy for halo mass in CDM haloes, which are described well by the Navarro--Frenk--White profile \citep[NFW;][]{NFW_96,NFW_97}: this profile can be collapsed into a one parameter model that depends only on halo mass, and its shape is altered little by stripping. However, the density profile of dark matter haloes changes significantly for dark matter models that are not CDM \citep[e.g.][Fig.~B1]{Lovell21} and therefore the interpretation of $V_\rmn{max}$ as a mass proxy in these models can be unclear. In this study we are able to test whether $V_\rmn{max}$ is still appropriate as a mass proxy for the novel vd100 model.

\begin{figure}
    \includegraphics[scale=0.22,angle=0]{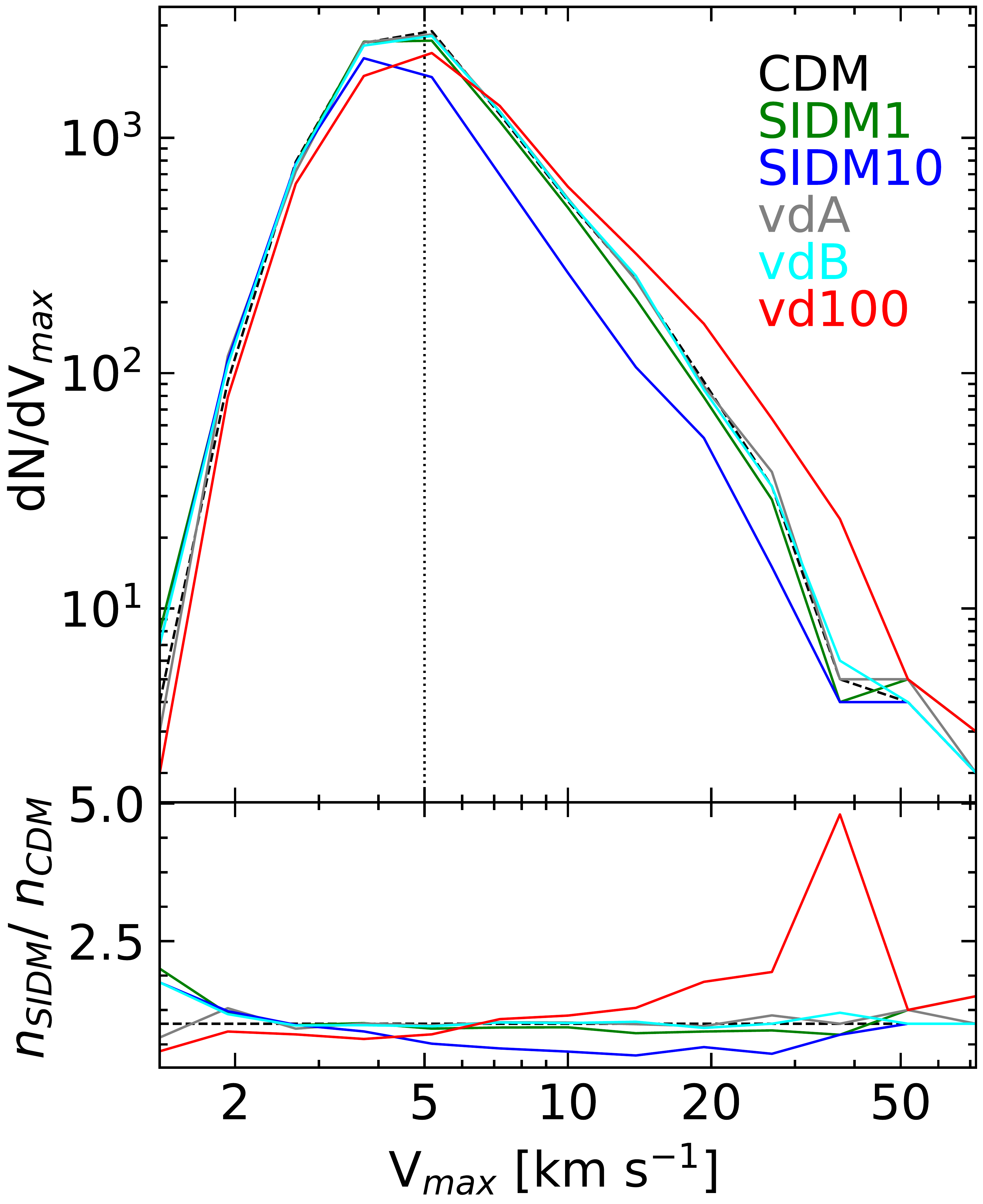}
    \caption{Differential subhalo $V_\rmn{max}$ function for all SIDM models considered, plus the CDM model. The The SIDM1, SIDM10, vd100, vdA and vdB models are denoted as green, blue, red, grey and cyan solid lines respectively. The CDM model is denoted by a black dashed line. The dotted vertical line shows the resolution limit of the simulations. The bottom panel shows the $V_\rmn{max}$ function of all SIDM models as a ratio to that of the CDM model.}
    \label{fig:velocity} 
\end{figure}

Fig.~\ref{fig:velocity} shows the differential subhalo maximum circular velocity function for all six simulations. As was previously shown in \citet{Vogelsberger12}, the SIDM1 model behaves quite similarly to the CDM model in terms of maximum circular velocity, as do the vdA and vdB models. The number of subhaloes in the SIDM10 model is significantly lower than that of the other dark matter models considered in this paper, again likely due to subhalo destruction from high velocity collisions. The most striking aspect of this figure is the very large number of vd100 subhaloes with $V_\rmn{max}>10\rmn{kms}^{-1}$: this model exhibits twice as many number of subhaloes with a $V_\rmn{max}$ of $\simeq$~20~km~s$^{-1}$ as the CDM simulation, despite there being marginally fewer subhaloes in total as shown in Fig.~\ref{fig:mass_1}. This phenomenon is a potential indicator as to the onset of gravothermal core collapse: in a fraction of the vd100 subhaloes, the inner regions of an initially cored subhalo collapse to form a cuspy profile. 

\begin{figure}
    \includegraphics[scale=0.42,angle=0]{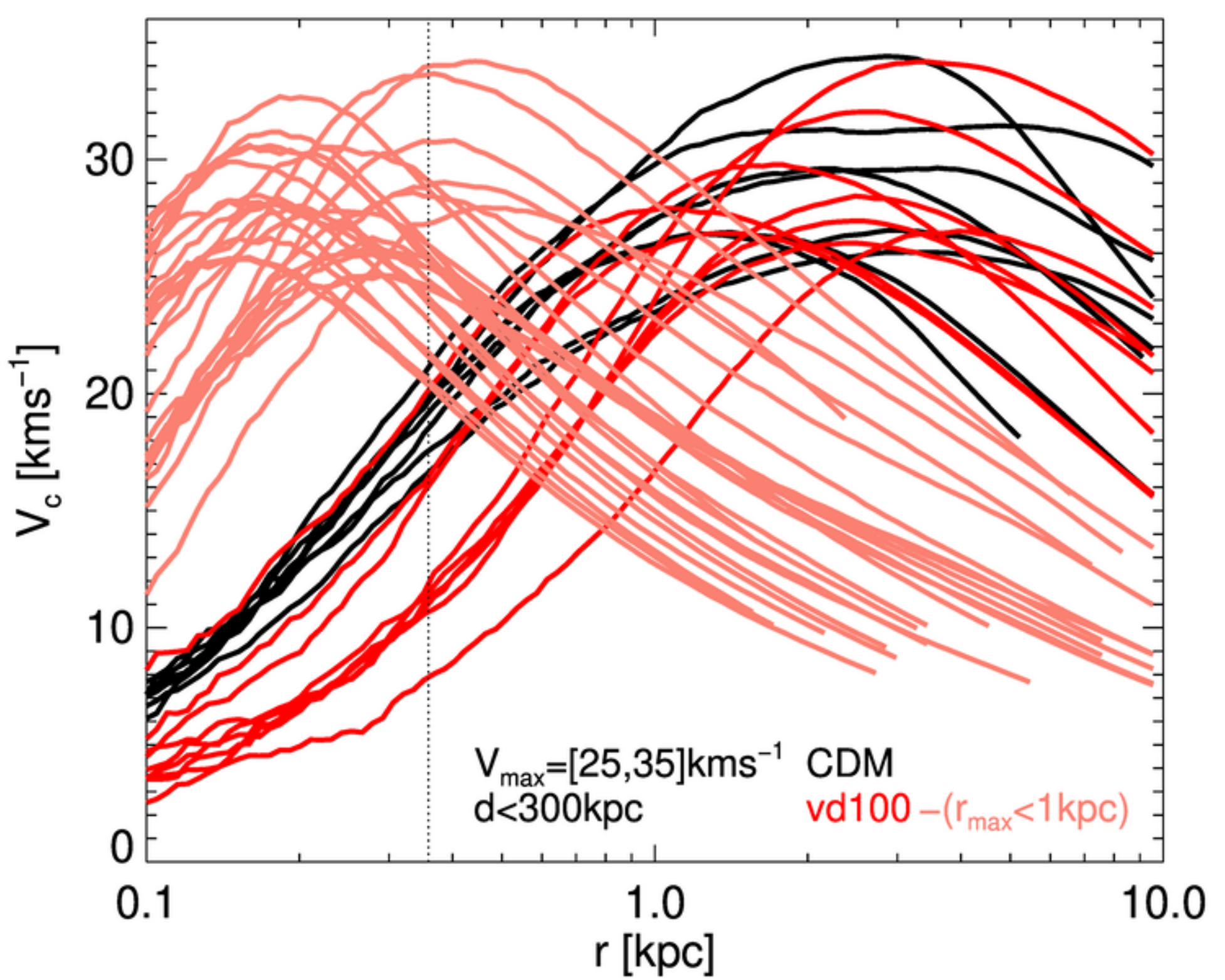}
    \caption{Circular velocity profiles of CDM and vd100 subhaloes at $z=0$. We select subhaloes that are within 300~kpc of the host centre and that have a $V_\rmn{max}$ value in the range 25~km~s$^{-1}$ to 35~km~s$^{-1}$. CDM curves are shown in black and vd100 in red. The vd100 curves for which $r_\rmn{max}$~<1~kpc have been faded to improve legibility. The dotted line denotes $2.8\times$ the gravitational softening length.}
    \label{fig:example_v_curve} 
\end{figure}

We demonstrate this phenomenon explicitly in Fig.~\ref{fig:example_v_curve}, which shows the circular velocity profiles of CDM and vd100 subhaloes at $z=0$. We select subhaloes that are within 300~kpc of the host centre and that have a $V_\rmn{max}$ value in the range 25~km~s$^{-1}$ to 35~km~s$^{-1}$. The number of CDM and vd100 subhaloes with $r_\rmn{max}$ $>1$~kpc in this range are approximately the same, with Fig.~\ref{fig:example_v_curve} showing 8 CDM subhaloes and 9 vd100 subhaloes. These regular vd100 haloes have cores and so are less dense than the CDM haloes but there is also a collapsed population present, where the circular velocity peak radius is significantly smaller than 1~kpc. In particular, $V_\rmn{max}$ can change drastically once the gravothermal collapse phase is triggered, since the collapse of the core makes the central regions even `hotter' dynamically and thus the value of $V_\rmn{max}$ increases (as can be inferred from Fig.~\ref{fig:example_v_curve}). There are 24 collapsed vd100 subhaloes, three times the number of uncollapsed subhaloes. We caution that the peaks of these collapsed curves lie within the gravitationally softened regime, and therefore the shapes of the circular velocity curves around the peak are highly uncertain.

\begin{figure*}
        \includegraphics[scale=0.279]{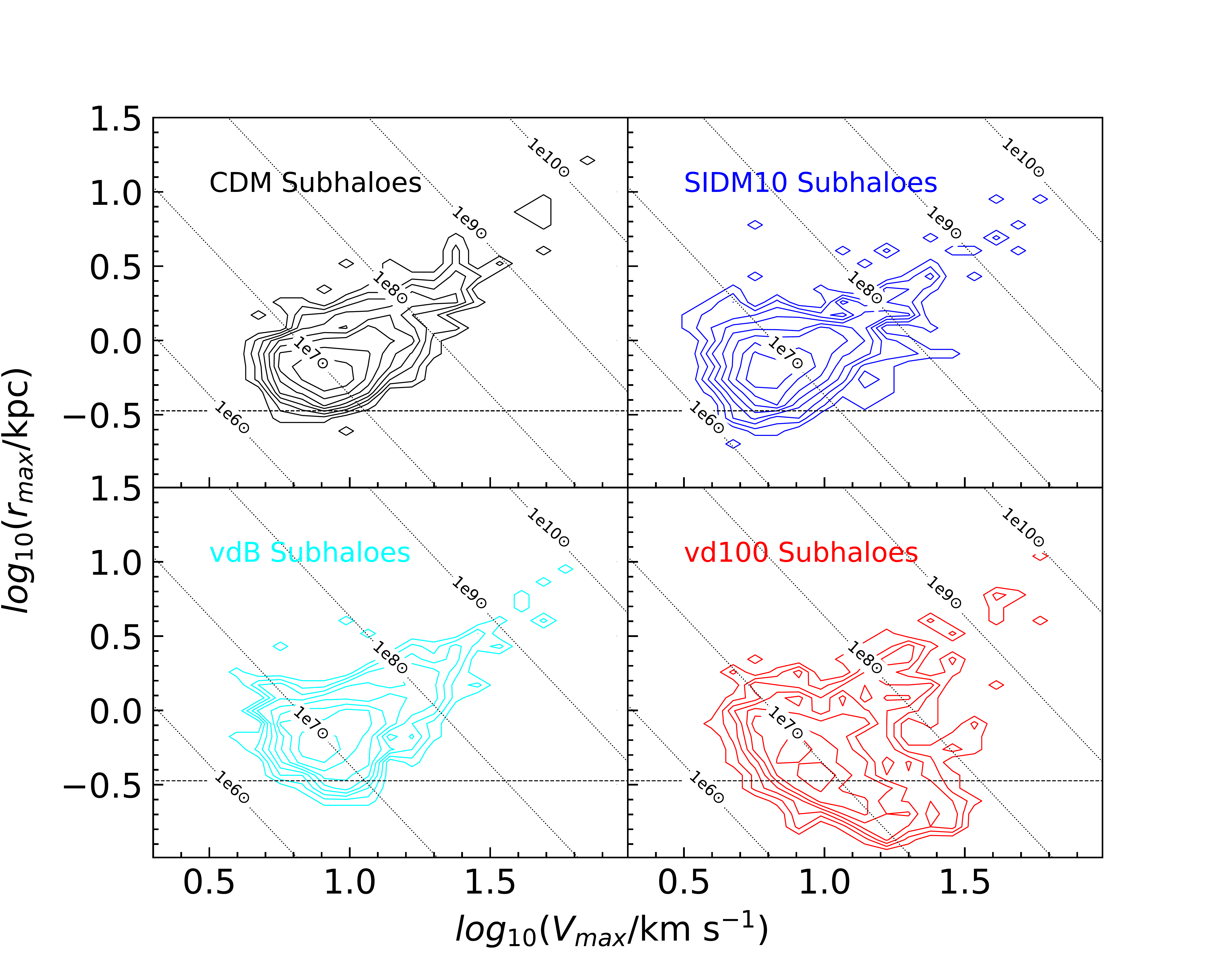}
    \caption{The $V_\rmn{max}$-$r_\rmn{max}$ 
    parameters for the CDM, SIDM10, vdB and vd100 subhalo samples, shown in the top left, top right, bottom left and bottom right panels respectively. The horizontal dashed line in each panel represents the 2.8~$\epsilon$ softening resolution limit. the dotted lines indicate the $V_\rmn{max}$--$r_\rmn{max}$ combinations that correspond to a fixed mass contained within $r_\rmn{max}$ as indicated in each panel.}
    \label{fig:V_Vs_R} 
\end{figure*}

In this situation, and in the context of equation~(\ref{eq:CircVel}), we anticipate a decrease in $r_\rmn{max}$ and an increase in $V_\rmn{max}$ at roughly constant $M_\rmn{sub}$. Therefore, we compute the distributions of $V_\rmn{max}$ and $r_\rmn{max}$ for the CDM, vdB, SIDM10 and vd100 models, and present the results for our simulations in Fig.~\ref{fig:V_Vs_R}; the SIDM1 and vdA versions of this plot are almost identical to the CDM version and are therefore omitted.

 The $V_\rmn{max}$-$r_\rmn{max}$ distributions for vdB and SIDM10 are broadly similar to that of CDM. SIDM10 exhibits marginally larger $r_\rmn{max}$ at fixed $V_\rmn{max}$ than is the case for CDM at the level of a few 10s of per~cent, which is consistent with the self-interactions carving out a core. vdB does not show a similar increase, and instead presents a small population of $V_\rmn{max}\sim10$~km~s$^{-1}$ haloes that have collapsed to $r_\rmn{max}$ values smaller than the softening. This trend increases dramatically for vd100 subhaloes: a large fraction of subhaloes in the mass range [$10^{7},10^{8}$]~$\msun$ have undergone collapse. We caution that many of these haloes have obtained values of $r_\rmn{max}$ that are smaller than the softening resolution limit. We are therefore able to show that collapse occurs and dramatically changes the density profile, but the accurate values of $V_\rmn{max}$ and $r_\rmn{max}$ are highly uncertain; we anticipate that the `true' $r_\rmn{max}$ would be smaller than we measure and thus $V_\rmn{max}$ even larger.  

This shift to a higher $V_\rmn{max}$ and lower $r_\rmn{max}$ is characteristic of gravothermal core collapse as the inner regions of the subhalo shrink in an accelerated collapse while retaining their mass. Fig.~\ref{fig:V_Vs_R} identifies the bound masses of the populations undergoing this gravothermal collapse in the vd100 model. It is evident that these are comparably lower to intermediate mass objects (within the range of approximately [$5\times10^{6}$, $1\times10^{8}$]~$\msun$.

\begin{figure*}
    \centering
    \includegraphics[scale=0.27]{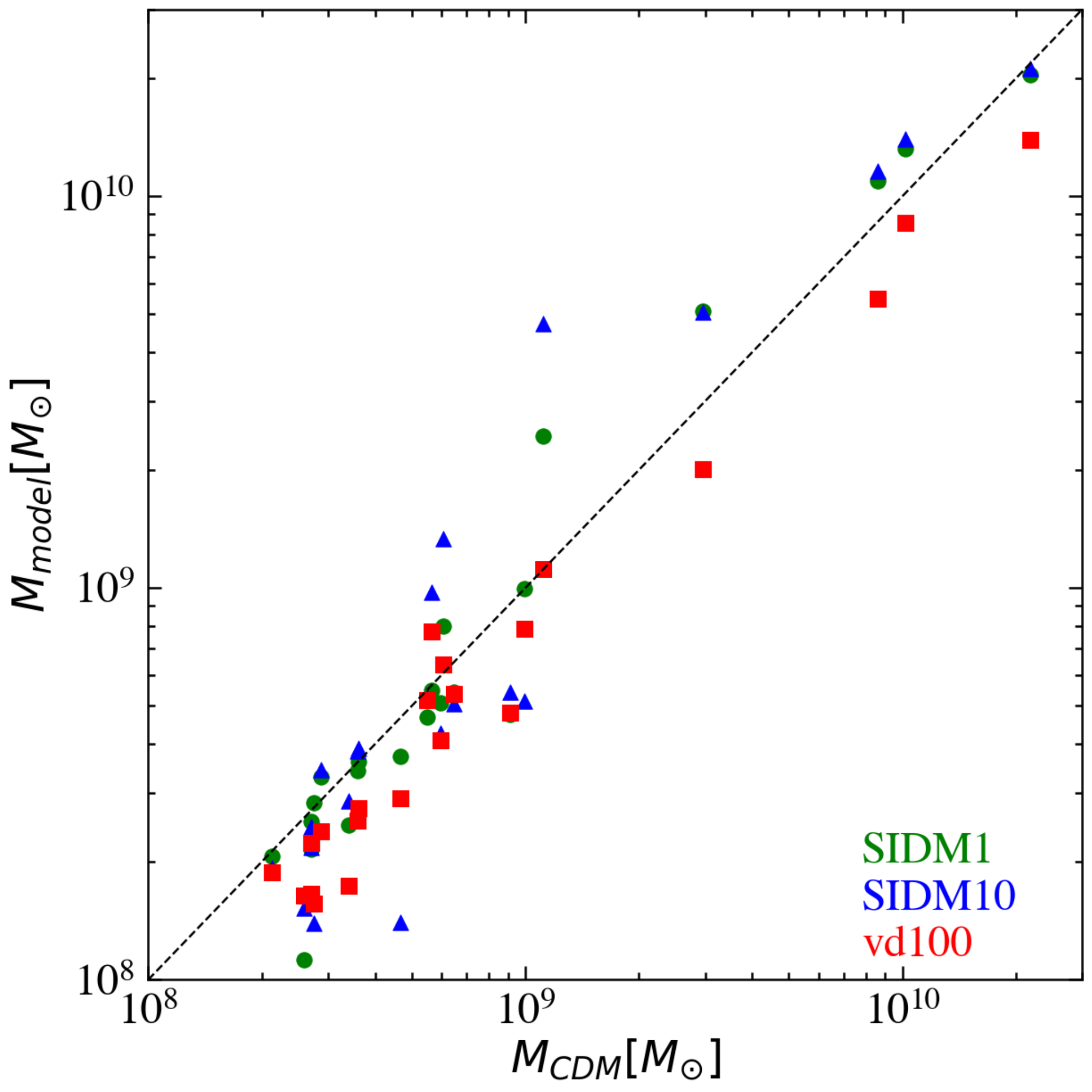}\hspace*{0.1cm}
    \includegraphics[scale=0.27]{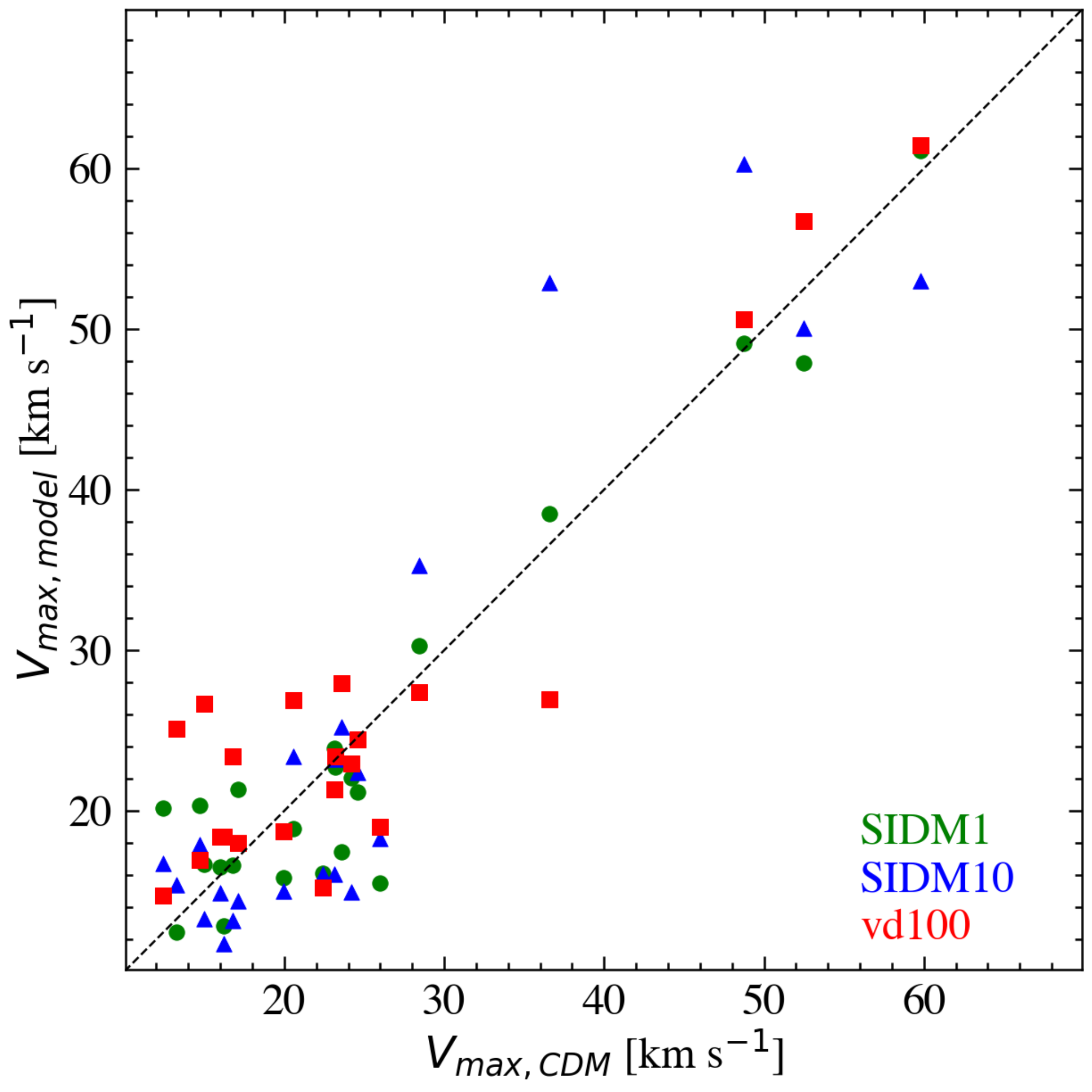}\hspace*{0.1cm}
    \includegraphics[scale=0.27]{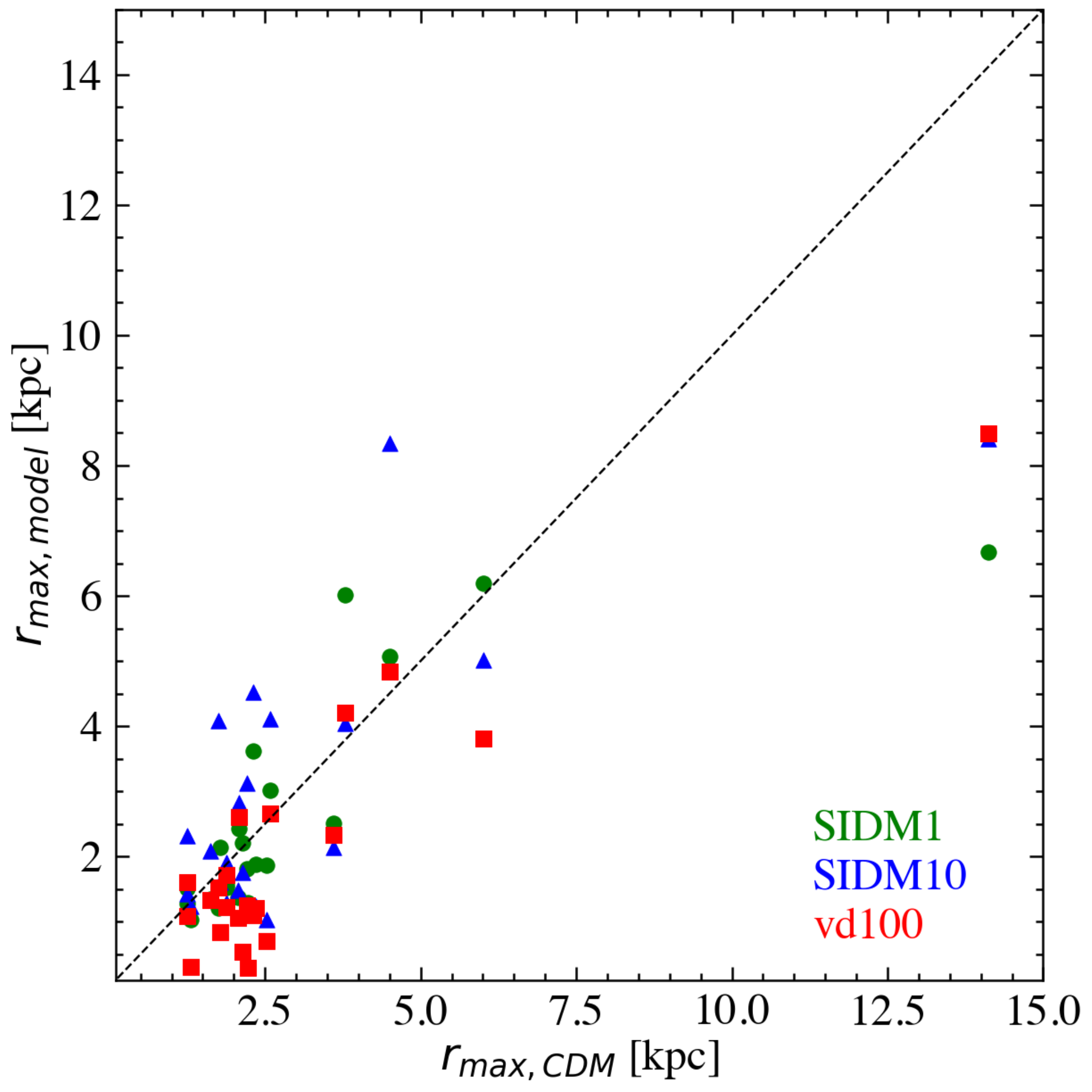}
    \caption{The properties of the same 22 subhaloes in the SIDM and CDM models matched between runs. In all three plots the vd100 model is denoted by red squares, the SIDM10 model by blue triangles and the SIDM1 model by green circles. Left: the bound mass of 22 matched subhaloes as simulated by the vd100 model and the SIDM models SIDM1 and SIDM10 shown in respect to the CDM model. Middle and right: the relationships between the aforementioned SIDM models and the CDM model in terms of $V_\rmn{max}$ and $r_\rmn{max}$ respectively.}
    \label{fig:matched1}
\end{figure*}

\subsubsection{Halo-by-halo}

We have shown that the vd100 model predicts a subhalo mass function that is only marginally suppressed relative to CDM, yet generates a novel population of high $V_\rmn{max}$ ($>10$~$\rmn{kms}^{-1}$) -- low $r_\rmn{max}$ ($<1$~kpc) subhaloes that are completely absent in CDM and the other SIDM models. We therefore hypothesise that a fraction of vd100 haloes undergo gravothermal collapse, retaining most of their mass while their $r_\rmn{max}$ shrinks dramatically. We examine this hypothesis further via two routes. First, we match a sample of subhaloes between vd100 and the other simulations, and second we consider how the properties of vd100 haloes change between $z=1$ and the present day.

We showed in Fig.~\ref{fig:HaloImagesHN} that the large scale structure and the positions of massive subhaloes are largely reproduced between the six models. The most divergent simulation in this regard is SIDM10, given the change in halo shape and the evaporation of massive haloes; never the less, we have used the Lagrangian halo matching method of \citet{Lovell14,Lovell18b} to identify 22 subhaloes that are present in four of our simulations: CDM, vd100, SIDM1, and SIDM10. Given the hypothesis outlined in the previous paragraph, we will determine whether the subhalo mass $M_\rmn{sub}$ is indeed constant between models, and how $V_\rmn{max}$ increases / $r_\rmn{max}$ decreases. We compare these three subhalo properties between simulations in Fig.~\ref{fig:matched1}, where we only consider haloes that have not been evaporated in any of the models. This could be biased to haloes that are on orbits with large pericentres or that have only recently fallen in, and have therefore not collapsed. We caution that the values of $r_\rmn{max}$ are not as well defined as those of $V_\rmn{max}$, and so could vary quite easily on a subhalo-by-subhalo basis even if the aggregate differences between the models are small.

In the left-hand panel, we show the bound mass of subhaloes in the SIDM1, SIDM10 and vd100 models in relation to the CDM model. There exists a population of subhaloes in the CDM subhalo mass range [$2\times10^{8}$,$5\times10^{8}$]$~\msun$ that are considerably less massive when simulated in vd100 than when modelled with CDM. We also find that nearly all subhaloes are $\sim$~30 per~cent less massive in the vd100 model than in the matched CDM subhalo, and indeed at higher masses are again much less massive than in all other considered models. This apparent decrease in bound mass of a large fraction of subhaloes in the vdSIDM model may be partially related to gravothermal core collapse in this selection of objects, as the rapid transfer of energy between inner and outer regions of each subhalo allows the outer components of the subhaloes to become unbound and so the total mass decreases. It is also possible that some of mass is removed by interactions with the host halo, although to a much less degree than for SIDM10.

In the middle panel of Fig.~\ref{fig:matched1}, we show the subhalo maximum circular velocities in the SIDM1, SIDM10 and vd100 models in relation to the CDM model, the latter of which sets a marker for what vd100 haloes could look like if they did not collapse. 13 of the 22 matched subhaloes display a higher value of $V_\rmn{max}$ when simulated in the vd100 model than in the CDM model and 6 of these subhaloes show $V_\rmn{max}$ values considerably greater in the vd100 model than in all others despite the fact that all of the vd100 subhaloes are less massive than their CDM counterparts. The right-hand panel of Fig.~\ref{fig:matched1} shows $r_\rmn{max}$ for the same subhaloes, and identifies a population of 5 subhaloes with an $r_\rmn{max}$ value less than 2.5kpc that, when modelled by the vd100 model, exhibit a significantly lower value of $r_\rmn{max}$ than in all other models. We therefore conclude that this population of high $V_\rmn{max}$ -- low $r_\rmn{max}$ vd100 haloes are collapsed versions of CDM haloes with $V_\rmn{max}<30$~$\rmn{kms}^{-1}$.

\begin{figure*}
\centering
\includegraphics[scale=0.2]{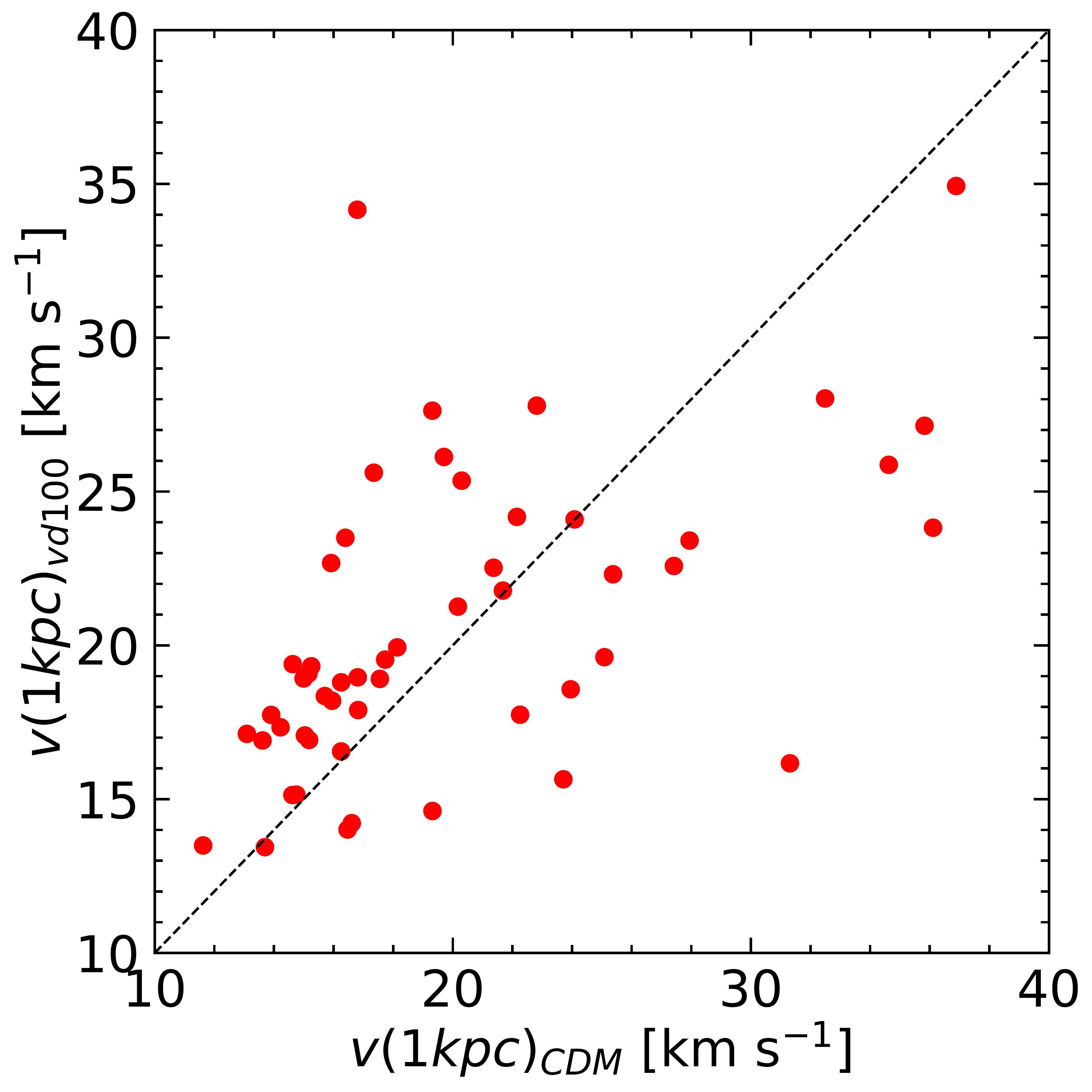}\hspace*{0.1cm}
\includegraphics[scale=0.2]{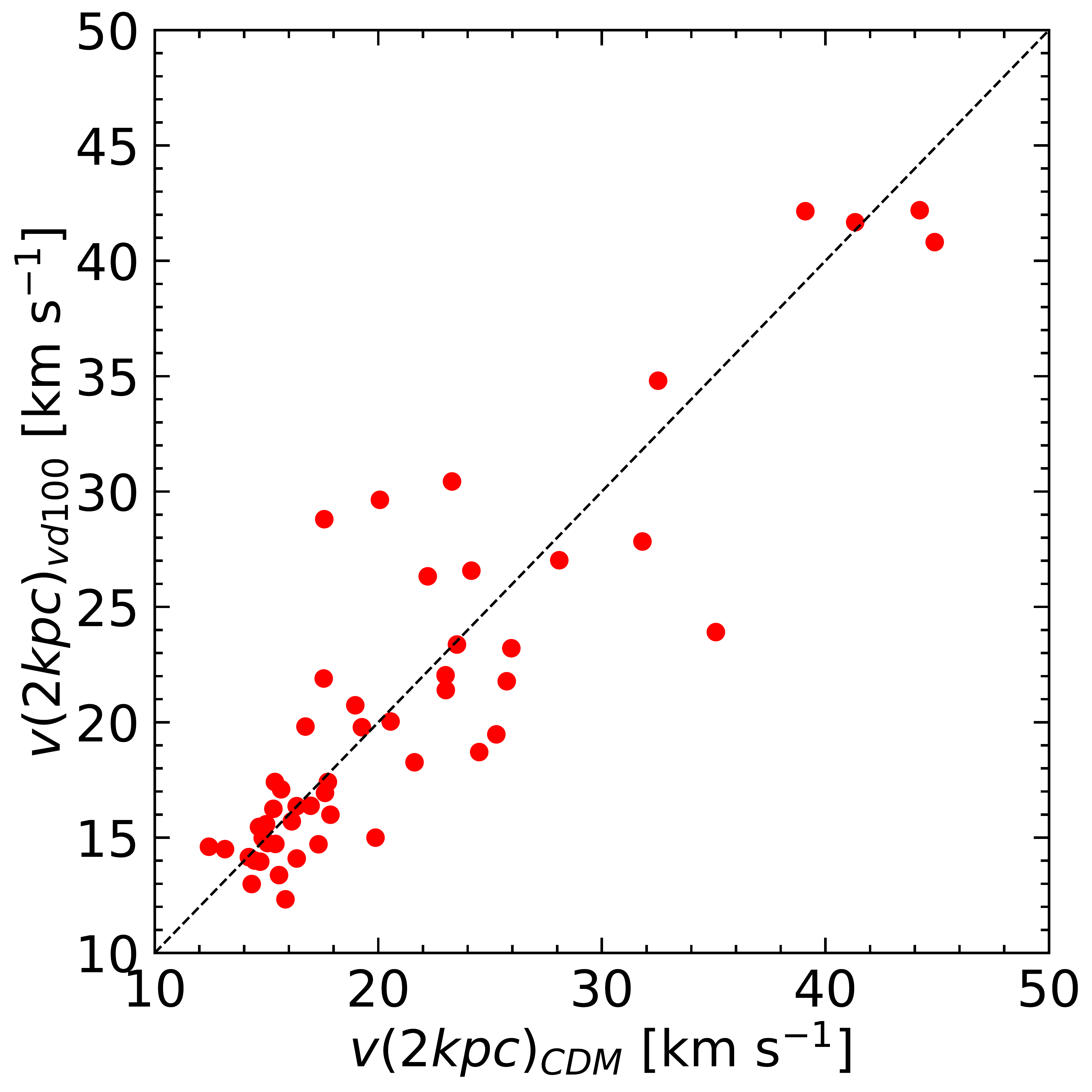}
\caption{The bound mass within 1~kpc and 2~kpc apertures (left and right panels respectively) of the centre of 63 matched subhaloes and the relationship between these values when modelled by the vd100 and CDM models.}
\label{fig:matched2}
\end{figure*}

This approach is somewhat complicated by the uncertainty in measuring $M_\rmn{sub}$, and also the problem that $r_\rmn{max}$ is only marginally resolved, if at all, in collapsed objects. We therefore take a complementary approach in which we instead measure the circular velocity within an aperture of fixed physical size. We choose two such apertures: 2~kpc and 1~kpc, and the circular velocities with these apertures are labelled v$_\rmn{1kpc}$ and v$_\rmn{2kpc}$. In the case that matter is ejected by self-interactions, the circular velocity -- which is directly related to the enclosed mass -- will decrease, and conversely if collapse from outside into the aperture then the circular velocity will instead increase. In order to obtain a larger sample of vd100 subhaloes, we generate a new list that does not require a match in SIDM1 or SIDM10; this approach returns 63 subhaloes. We plot the vd100-CDM matched pair circular velocities at 1 and 2~kpc apertures in Fig.~\ref{fig:matched2}.

In the 1 kpc aperture there is a significant fraction of subhaloes that have a higher v$_\rmn{1kpc}$ in the vd100 model than in the CDM model, whilst a smaller number of the CDM subhaloes with v$_\rmn{1kpc}>22$~kms$^{-1}$ show a reduced mass in this aperture under vdSIDM constraints relative to CDM. This result is consistent with the picture that the more massive vd100 objects have had a core scooped out by self-interactions whereas in the lower mass haloes sufficient time has elapsed for that core to collapse, with material from outside the aperture flowing towards the subhalo centre. An interesting corollary of this result is that the vd100 haloes have a much narrower range of central masses than do their CDM counterparts: in this extreme vd100 model, we therefore predict that the MW satellites should inhabit a narrow range of masses, corresponding to 1~kpc ciruclar velocities of [15-30]~km~s$^{-1}$. The right-hand panel of Fig.~\ref{fig:matched2} instead uses an aperture of 2~kpc, within which there is scatter about the 1:1 relation. We therefore conclude that the novel behaviour of this model is largely confined to within 1~kpc of the subhalo centre.

\begin{figure}
    \includegraphics[scale=0.219,angle=0]{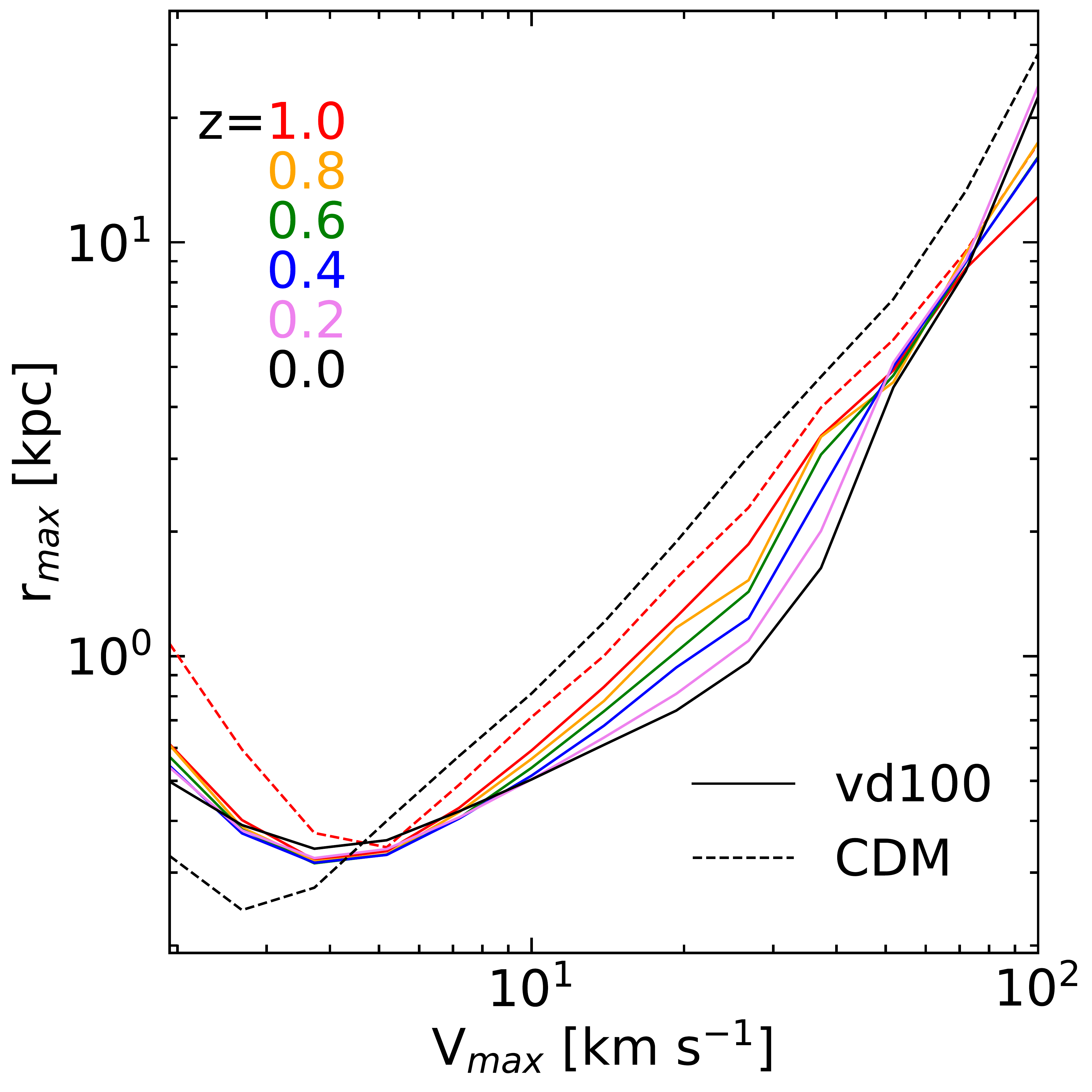}
    \caption{The evolution of the simulated vd100 dark matter subhalo $V_\rmn{max}$-$r_\rmn{max}$ parameters over time. The circular velocity profile for the same subhaloes modelled under the CDM regime can also be seen at redshift values of $z=1$ and $z=0$ for comparison, plotted as dashed lines. The colours plotted correspond to a redshift of $z=1$ (red), $z=0.8$ (yellow), $z=0.6$ (green), $z=0.4$ (blue), $z=0.2$ (violet) and $z=0$ (black).}
    \label{fig:Evolution} 
\end{figure}

\subsection{Time evolution of the subhalo mass profile}

We have shown thus far that there exists a large population of vd100 subhaloes that are much more concentrated than CDM subhaloes, and that at least some of these subhaloes correspond to CDM haloes where the mass has become concentrated within 1~kpc. From hereon in we present the time evolution of the vd100 subhaloes, first in the population as a whole and subsequently for a subsample of subhaloes. First, in Fig.~\ref{fig:Evolution}, we track the evolution of the mean $V_\rmn{max}$-$r_\rmn{max}$ relation for vdSIDM subhaloes from $z=1$ to $z=0$.

The subhaloes in the vd100 model show $V_\rmn{max}$-$r_\rmn{max}$ parameters comparable to that of those in the CDM model at a redshift of $z=1$, but at later times there is a steady preference for $r_\rmn{max}$ to decline at fixed $V_\rmn{max}$. The most dramatic of these departures occurs between $z=1$ and $z=0.8$ for subhaloes with a maximum circular velocity that lies between 20 and 30~km\,s$^{-1}$. This trend is the reverse of CDM, in which the population of haloes of a given mass / $V_\rmn{max}$ becomes less concentrated with time and therefore the mean $r_\rmn{max}$ increases from $z=1$ to $z=0$.

\begin{figure*}
    \includegraphics[scale=0.21,angle=0]{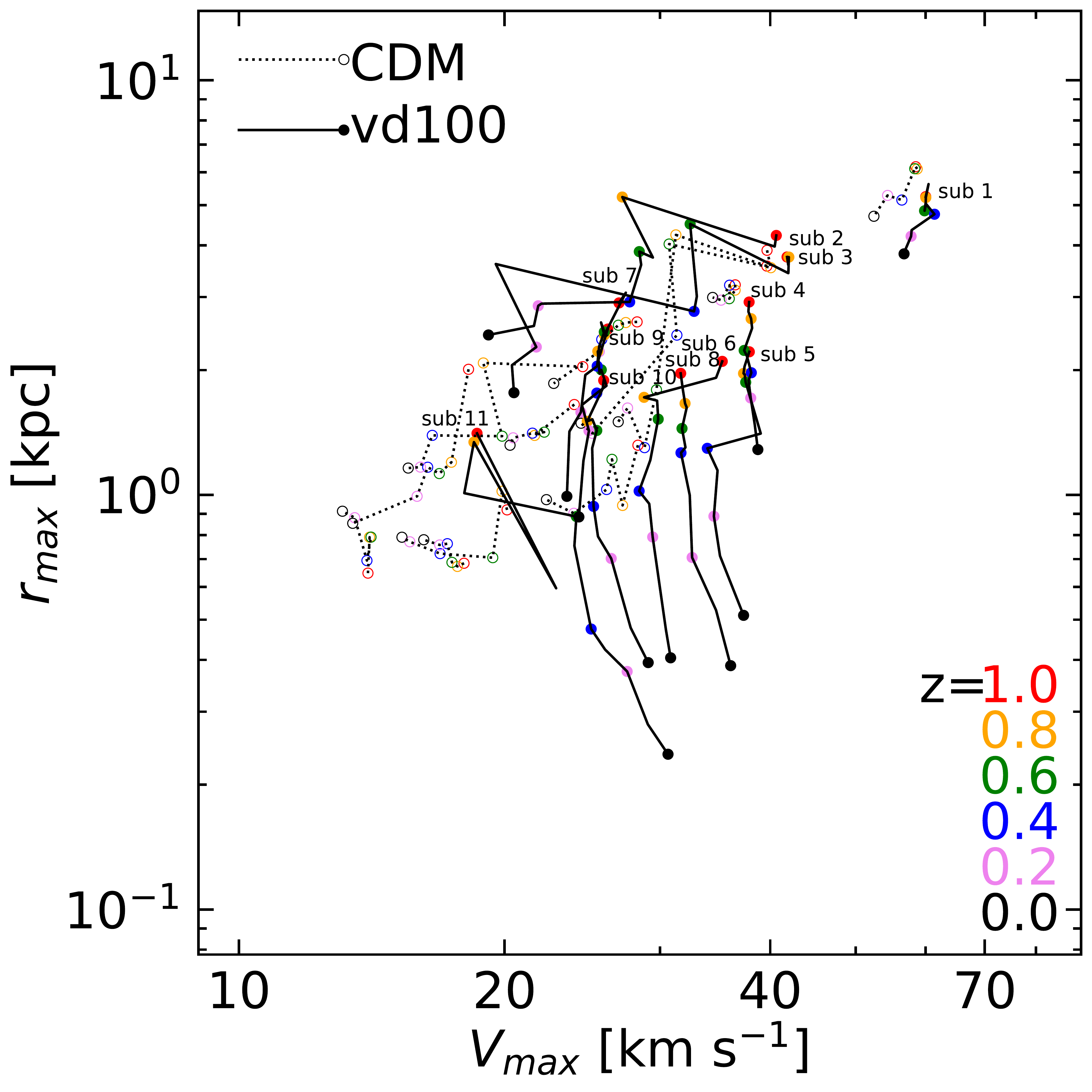}\hspace*{0.2cm}
    \includegraphics[scale=0.21,angle=0]{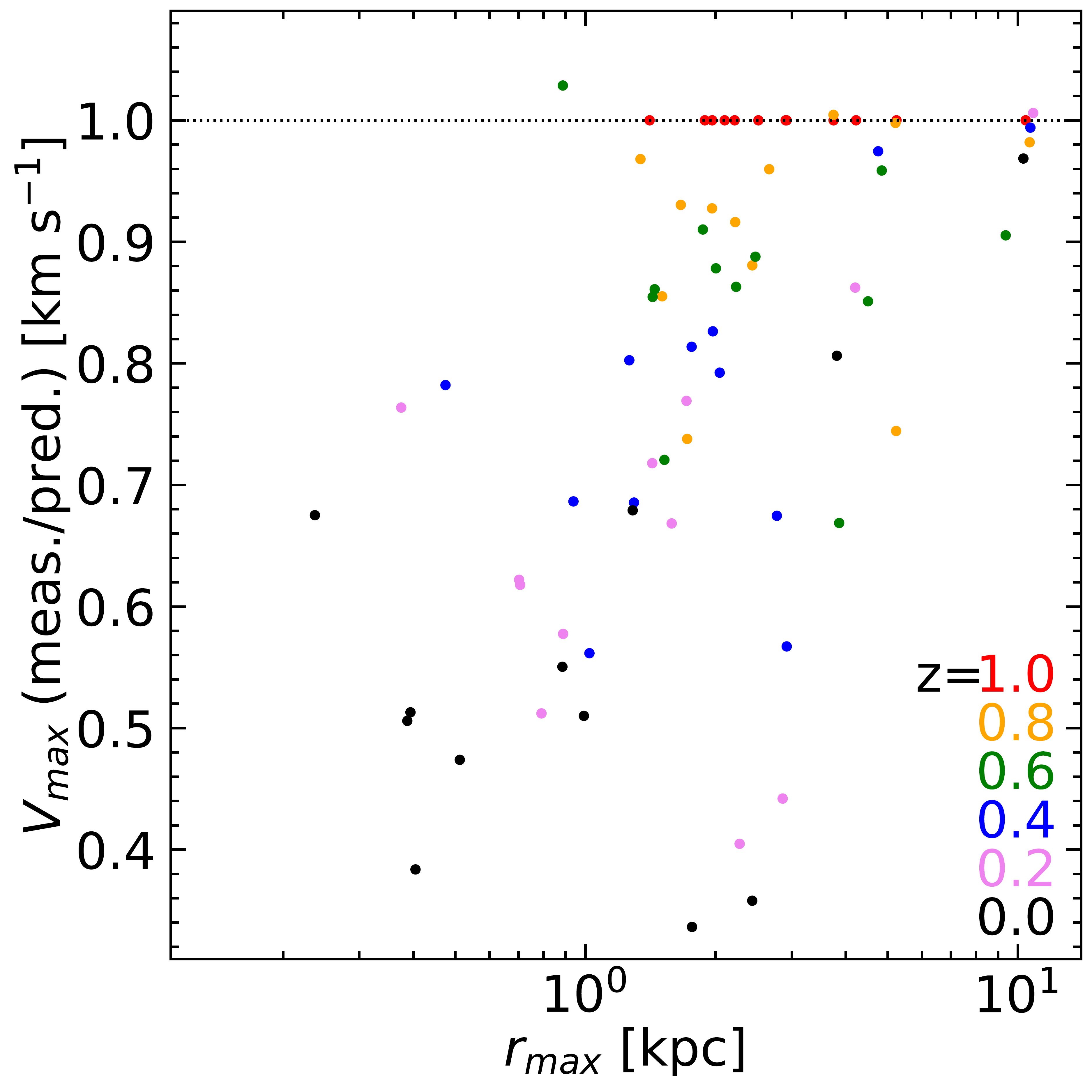}
    \caption{Left-hand panel: The evolution of the $V_\rmn{max}$--$r_\rmn{max}$ 
    parameters from $z=1$ to $z=0$ for 11 vd100 and 11 CDM subhaloes. The coloured dots represent the $V_\rmn{max}$-$r_\rmn{max}$ parameter for the subhalo at the given redshift, the black lines trace the evolutionary track for the individual halo, moving from red ($z=1$) to black ($z=0$). The CDM data are shown with empty circles and dotted lines, and vd100 with filled circles as solid lines. Right-hand panel: the ratio of $V_\rmn{max}$ (measured) to $V_\rmn{max}$ (predicted) for the vd100 model, shown as a function of the corresponding $r_\rmn{max}$ for 11 individual subhaloes from $z=1$ to $z=0$.}
    \label{fig:MatchedZTrack} 
\end{figure*}

The path of individual haloes through the $V_\rmn{max}$-$r_\rmn{max}$ diagram cannot be discerned directly from Fig.~\ref{fig:Evolution}. In order to find the direction in which haloes move from the CDM-like mean $V_\rmn{max}$-$r_\rmn{max}$ relation into the population of highly concentrated vd100 subhaloes, we identify a subsample of 11 subhaloes, plus the host halo, at the 11 output times that we have available. We then plot the evolution tracks of these objects through the $V_\rmn{max}$-$r_\rmn{max}$ in the redshift interval $[1,0]$ in  Fig.~\ref{fig:MatchedZTrack}. We also include 11 CDM subhaloes that span the $z=1$ $V_\rmn{max}$ range of the vd100 subsample.

All of the subhaloes bar 2 and 3 evolve towards a lower value of $r_\rmn{max}$ and, excluding subhalo 11, show an initial decrement in $V_\rmn{max}$ which nevertheless increases again slightly by $z=0$. Subhaloes 2 and 3 are outliers in that their $V_\rmn{max}$ instead consistently decrease in value roughly in line with the mean $z=1$ $V_\rmn{max}$-$r_\rmn{max}$ relation, therefore show similar tracks to the CDM subhalo evolution. 

If these subhaloes were acting purely under the influence of collapse, with all of the mass within the $z=1$ $r_\rmn{max}$ retained to $z=0$, these subhaloes would have shifted to higher $V_\rmn{max}$ than we found in this simulation, and therefore it would appear that tidal stripping has taken place. We check the degree to which this is the case by computing, for each of these subhaloes, the mass within $r_\rmn{max}$ at $z=1$, and then calculating the $V_\rmn{max}$ that each subhalo would have had if the mass enclosed within $r_\rmn{max}$ at each subsequent snapshot were the same as we calculated at $z=1$. We compute the ratio of the measured $V_\rmn{max}$ to this `predicted' $V_\rmn{max}$ and plot the results as a function of $r_\rmn{max}$ in the right-hand panel of  Fig.~\ref{fig:MatchedZTrack}.

With the exception of a single subhalo at a redshift of $z=0.6$, all considered subhaloes lose mass between $z=1$ and $z=0$. The mass loss is between 20~per~cent and 65~per~cent, which we expect is due to tidal stripping rather than to the change of the dark matter density profile. 

\begin{figure*}
    \includegraphics[scale=0.65,angle=0]{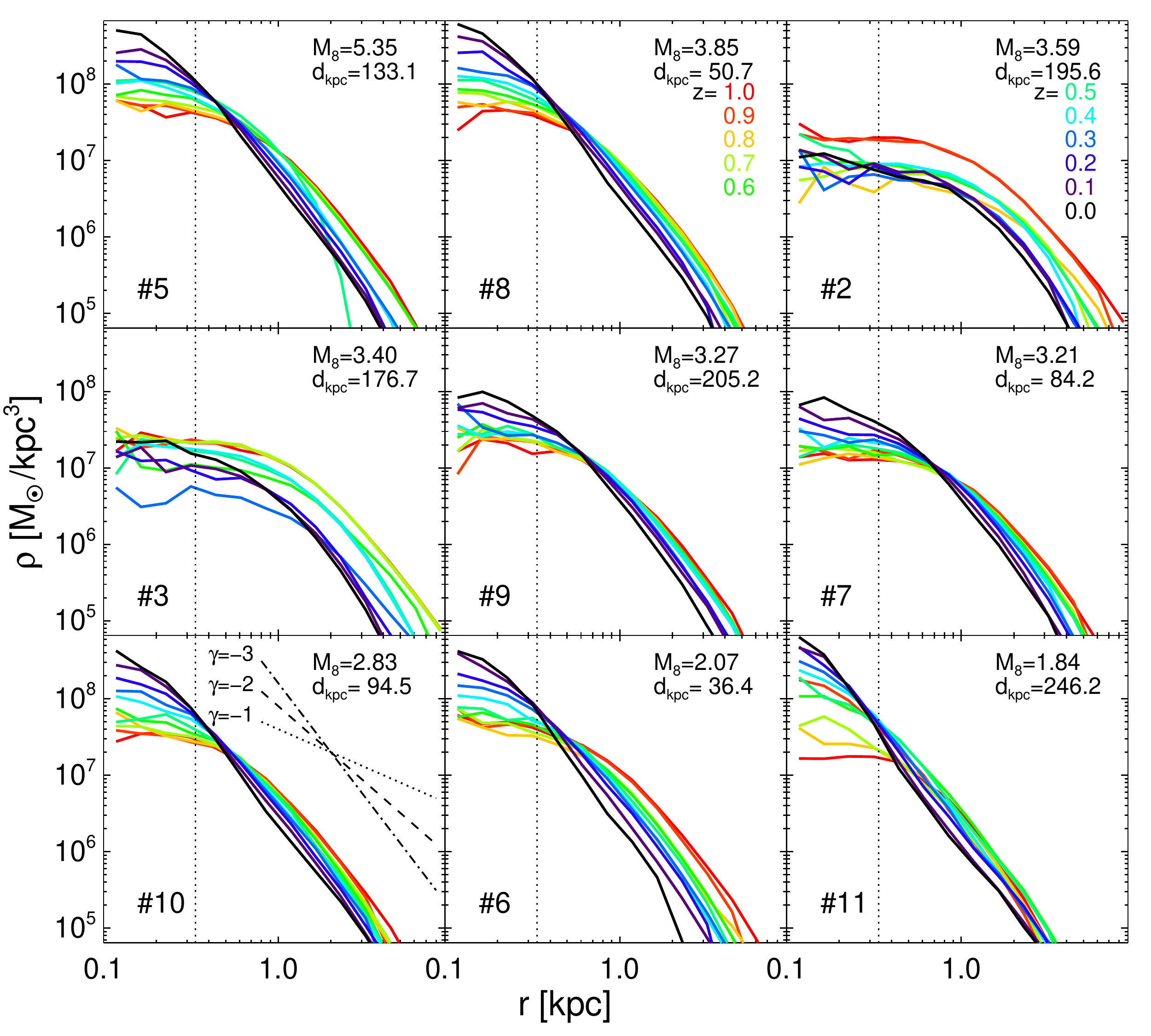}
    \caption{Density profiles for nine of the 11 haloes presented in Fig.~\ref{fig:MatchedZTrack}, with the correspondence indicated by the number in the bottom left hand corner of each panel; we also include the $z=0$ subhalo mass in units of $10^{8}\msun$ and the $z=0$ distance from the main halo centre. We show density profiles at 11 redshifts, which are shown with different colours evolving from red at $z=1$ to black at $z=0$. The dotted vertical lines indicate the radius equal to 2.8 times the softening length, and therefore mark the hard resolution limit of our simulations. In the bottom left-hand panel we indicate power law slopes with indicies -1, -2, and -3 as dotted, dashed, and dot-dashed lines respectively.}
    \label{fig:ExampDP} 
\end{figure*}

We end our presentation of the results by showing explicitly how the matter distribution changes between $z=1$ and $z=0$ for nine of the vd100 haloes discussed in Fig.~\ref{fig:MatchedZTrack}. First, we present the density profiles of these nine subhaloes in Fig.~\ref{fig:ExampDP}, with one curve per redshift, i.e. per output time. 

\begin{figure*}
    \includegraphics[scale=0.65,angle=0]{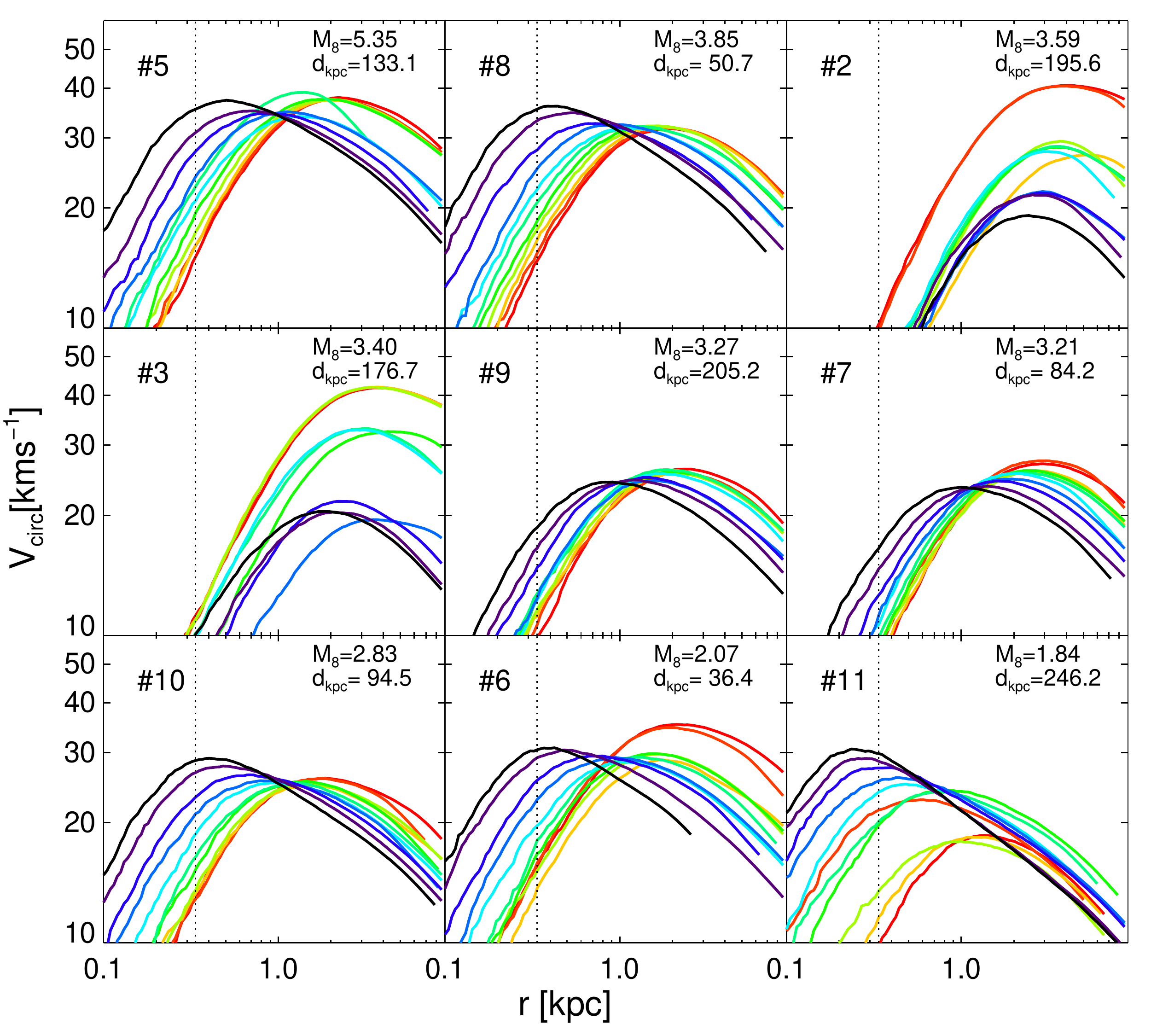}
    \caption{Same as Fig.~\ref{fig:ExampDP} but for the circular velocity profiles.}
    \label{fig:ExampVP} 
\end{figure*}

Three of the nine subhaloes -- 2, 3, and 7 -- all show classic SIDM cored profiles at $z=1$. This profile persists to $z=0$ for subhalo 2, albeit with plenty of stripping, and subhalo 3 is similar. The other haloes instead show very different results: their $z=1$ profiles are slightly steeper than for the two subhaloes mentioned previously, and over the rest of the simulation they evolve to a very cuspy profile with a power law index of $\gamma$~$\approx-3$; this change in density profile is again indicative of gravothermal collapse. Given that the $z=1$ profiles are not as cored as haloes 2 and 3, we suspect that this process of collapse may have begun at some earlier time; we are unable to check whether this is the case because snapshot outputs were only stored from $z=1$ onwards. There is some evidence that subhaloes that are relatively close to the halo centre at $z=0$ have shown a sharper collapse, but on the other hand subhaloes 7 and 9 exhibit very similar profiles despite differing in distance by a factor of two. We therefore do not have strong evidence at this stage to support the claim that gravothermal collapse is induced/enhanced by tidal interactions: ultimately a detailed study of the subhalo orbits will be required to answer this question.  

Finally, in Fig.~\ref{fig:ExampVP} we present the circular velocity profiles for the nine haloes from Fig.~\ref{fig:ExampDP}. For most of the subhaloes shown, the maximum circular velocity ($V_\rmn{max}$) remains almost constant, to within a few tens of per~cent, whilst the corresponding radius ($r_\rmn{max}$) decreases from a redshift of $z=1$ to $z=0$. This is again with the exception of subhaloes 2 and 3, which evolve to have a circular velocity profile with a considerably lower $V_\rmn{max}$ and a slightly decreased $r_\rmn{max}$, and this likely due to tidal stripping as stated above. Subhalo 11 is also an exception as it evolves to have a considerably increased maximum circular velocity. We also see how the $r_\rmn{max}$ value for subhalo 11 decreases dramatically between $z=1$ and $z=0.9$ and then increases again at $z=0.8$, which is unlike the other subhaloes shown which demonstrate a fairly sequential decrease in $r_\rmn{max}$. We caution that for the collapsed objects,  $r_\rmn{max}$ occurs around the hard resolution limit of $2.8\times$ the Plummer radius. It is therefore well within the Power radius \citep{Power03,Springel08b} that is typically assumed to provide a rigorous limit on the resolution of the density profile in CDM $N$-body simulations. It is not clear that this limit is appropriate for these simulations: we simply state that the interpretation for these simulations for the purpose of comparison must be done with care, as it is not clear whether these circular velocities are completely free of numerical artifacts around the measured $r_\rmn{max}$, although we not note that the overall behaviour does follow the general expectations for the gravothermal collapse.

\section{Conclusions}
\label{sec:con}

The relationship between the masses of Milky Way (MW) satellite galaxies on the one hand and the nature of the dark matter on the other remains unclear, despite a plethora of explanations based on baryon physics \citep{NEF96,diCintio11,Pontzen_Governato_11,Sawala16a} and modifications to the dark matter physics \citep{Lovell12,Vogelsberger12,Zavala13,CyrRacine16,Schewtschenko16,Vogelsberger16}. \citet{Zavala2019} showed that models of self-interacting dark matter (SIDM) can explain the full set of known satellite masses if the self-interaction cross-section at low relative velocities is 100~$\rmn{cm}^{2}\rmn{g}^{-1}$, likely through the gravothermal collapse of a fraction of subhaloes to form very dense subhaloes. In this paper we have studied the possibility for this collapse in detail.

We have used $N$-body simulations of a MW-analogue halo to analyse and compare SIDM models of both the velocity independent and velocity dependent varieties, and contrasted them with the cold dark matter (CDM) model. We focus particularly on the extreme 100~$\rmn{cm}^{2}\rmn{g}^{-1}$ velocity independent SIDM model (vdSIDM) model run by \citet{Zavala2019}, which we refer to as vd100. We also used four simulations run by \citep{Vogelsberger12} and \citep{Zavala13}, two velocity independent ($1$~$\rmn{cm}^{2}\rmn{g}^{-1}$ and $10$~$\rmn{cm}^{2}\rmn{g}^{-1}$) and two velocity dependent, vdA and vdB (see Table~\ref{tab:table} for their descriptions). 

The subhalo mass function (Fig.~\ref{fig:mass_1}) differs only slightly between CDM and vdSIDM models. What suppression there is relative to the CDM model comes at $M_\rmn{sub}<10^{8}\msun$, and these differences do not become much stronger with distance to the host halo centre (Fig.~\ref{fig:mass_rad}). However, pronounced differences between vd100 and all of the other models occur when we instead consider the mass profile of the subhaloes (Fig.~\ref{fig:velocity}). Our main conclusions are as follows:

\begin{enumerate}[leftmargin=1.099cm, label=(\roman*)]  
    \item The vd100 halo mass function is suppressed at the 10~per~cent level for $M_\rmn{sub}<10^{8}$~$\msun$ relative to CDM scenarios.

    \item  The vd100 model hosts a significantly greater number of subhaloes with a higher value of maximum circular velocity ($V_\rmn{max}$~$\geq$~10~km~s$^{-1}$) than in the CDM model (Fig.~\ref{fig:velocity}). These subhaloes are part of the subhalo population that have values of $r_\rmn{max}$ much smaller than is the case for either CDM or for other SIDM models (Fig.~\ref{fig:V_Vs_R}). These properties are characteristic of gravothermal core collapse as subhaloes that initially displayed a cored density profile collapse to a cuspy profile. It is evident through subhalo tracking that a fraction of individual subhaloes in the vd100 model that had $V_\rmn{max}$-$r_\rmn{max}$ parameters typical of CDM at $z=1$, evolve to very low $r_\rmn{max}$ -- as low as the spatial resolution of the simulation -- by $z=0$ while roughly retaining their $z=1$ $V_\rmn{max}$ value (Fig.~\ref{fig:Evolution}).  
    
    \item Populations undergoing gravothermal core collapse are shown to be the lower to intermediate mass subhaloes in the simulation, all within the mass range of $5\times10^{6}\leq M_\rmn{sub}\leq$~$1\times10^{8}$~$\msun$ (Fig.~\ref{fig:V_Vs_R}). 

\end{enumerate}

We argue that our results demonstrate clear evidence of gravothermal core collapse occurring in a fraction of our simulated dark matter subhaloes between redshifts of $z=1$ and $z=0$. The inclusion of a velocity dependence on the self-scattering cross-section in the vdSIDM model allows higher values of cross-sections to be explored at low velocities, enabling an increased probability of gravothermal collapse and leading to an appreciable diversity in circular velocity profiles in our simulated subhaloes, as is in accordance with the findings of \citet{Zavala2019}. 

Whilst beyond the scope of this paper, we note that the impact of baryonic matter in a full hydrodynamic simulation of vdSIDM is crucial for the evolution of dark matter subhaloes with a large self-scattering cross-section and the onset of gravothermal collapse as changes in the baryonic potential has been shown to accelerate the expansion and subsequent contraction of the SIDM core \citep[see][]{Sameie18}, similar to the effects we see in our simulation due to tidal stripping. We also propose that consideration of the initial properties of the subhaloes undergoing gravothermal collapse may prove an interesting area for further research.  

\section*{Acknowledgements}
HT acknowledges the support and generosity of the University of Hull Alumni and their donations aimed at widening participation and gender diversity, alongside the Student Futures University of Hull Internship. MRL and JZ acknowledge support by a Grant of Excellence from the Icelandic Research Fund (grant number 173929). MV acknowledges support through a NASA ATP grant NNX17AG29G, and NSF grants AST-1814053, AST-1814259,  AST-1909831 and AST-2007355.

\section*{Data availability statement}
 The simulation data were first generated for the papers outline in Table~\ref{tab:table}.  Researchers wishing to gain access to these data should contact the lead author of each of those papers.





\bibliographystyle{mnras}

\begin{thebibliography}{}
\makeatletter
\relax
\def\mn@urlcharsother{\let\do\@makeother \do\$\do\&\do\#\do\^\do\_\do\%\do\~}
\def\mn@doi{\begingroup\mn@urlcharsother \@ifnextchar [ {\mn@doi@}
  {\mn@doi@[]}}
\def\mn@doi@[#1]#2{\def\@tempa{#1}\ifx\@tempa\@empty \href
  {http://dx.doi.org/#2} {doi:#2}\else \href {http://dx.doi.org/#2} {#1}\fi
  \endgroup}
\def\mn@eprint#1#2{\mn@eprint@#1:#2::\@nil}
\def\mn@eprint@arXiv#1{\href {http://arxiv.org/abs/#1} {{\tt arXiv:#1}}}
\def\mn@eprint@dblp#1{\href {http://dblp.uni-trier.de/rec/bibtex/#1.xml}
  {dblp:#1}}
\def\mn@eprint@#1:#2:#3:#4\@nil{\def\@tempa {#1}\def\@tempb {#2}\def\@tempc
  {#3}\ifx \@tempc \@empty \let \@tempc \@tempb \let \@tempb \@tempa \fi \ifx
  \@tempb \@empty \def\@tempb {arXiv}\fi \@ifundefined
  {mn@eprint@\@tempb}{\@tempb:\@tempc}{\expandafter \expandafter \csname
  mn@eprint@\@tempb\endcsname \expandafter{\@tempc}}}

\bibitem[\protect\citeauthoryear{{Balberg} \& {Shapiro}}{{Balberg} \&
  {Shapiro}}{2002}]{Balberg02}
{Balberg} S.,  {Shapiro} S.~L.,  2002, \mn@doi [\prl]
  {10.1103/PhysRevLett.88.101301}, \href
  {https://ui.adsabs.harvard.edu/abs/2002PhRvL..88j1301B} {88, 101301}

\bibitem[\protect\citeauthoryear{{Boylan-Kolchin}, {Bullock}  \&
  {Kaplinghat}}{{Boylan-Kolchin} et~al.}{2011}]{BoylanKolchin11}
{Boylan-Kolchin} M.,  {Bullock} J.~S.,   {Kaplinghat} M.,  2011, \mn@doi
  [\mnras] {10.1111/j.1745-3933.2011.01074.x}, \href
  {http://adsabs.harvard.edu/abs/2011MNRAS.415L..40B} {415, L40}

\bibitem[\protect\citeauthoryear{{Boylan-Kolchin}, {Bullock}  \&
  {Kaplinghat}}{{Boylan-Kolchin} et~al.}{2012}]{BoylanKolchin12}
{Boylan-Kolchin} M.,  {Bullock} J.~S.,   {Kaplinghat} M.,  2012, \mn@doi
  [\mnras] {10.1111/j.1365-2966.2012.20695.x}, \href
  {http://adsabs.harvard.edu/abs/2012MNRAS.422.1203B} {422, 1203}

\bibitem[\protect\citeauthoryear{{Bullock} \& {Boylan-Kolchin}}{{Bullock} \&
  {Boylan-Kolchin}}{2017}]{bullock2017}
{Bullock} J.~S.,  {Boylan-Kolchin} M.,  2017, \mn@doi [\araa]
  {10.1146/annurev-astro-091916-055313}, 55, 343

\bibitem[\protect\citeauthoryear{{Col{\'{\i}}n}, {Avila-Reese}, {Valenzuela}
  \& {Firmani}}{{Col{\'{\i}}n} et~al.}{2002}]{Colin02}
{Col{\'{\i}}n} P.,  {Avila-Reese} V.,  {Valenzuela} O.,   {Firmani} C.,  2002,
  \mn@doi [\apj] {10.1086/344259}, \href
  {http://adsabs.harvard.edu/abs/2002ApJ...581..777C} {581, 777}

\bibitem[\protect\citeauthoryear{{Correa}}{{Correa}}{2020}]{Correa20}
{Correa} C.~A.,  2020, arXiv e-prints, \href
  {https://ui.adsabs.harvard.edu/abs/2020arXiv200702958C} {p. arXiv:2007.02958}

\bibitem[\protect\citeauthoryear{{Cyr-Racine}, {Sigurdson}, {Zavala},
  {Bringmann}, {Vogelsberger}  \& {Pfrommer}}{{Cyr-Racine}
  et~al.}{2016}]{CyrRacine16}
{Cyr-Racine} F.-Y.,  {Sigurdson} K.,  {Zavala} J.,  {Bringmann} T.,
  {Vogelsberger} M.,   {Pfrommer} C.,  2016, \mn@doi [\prd]
  {10.1103/PhysRevD.93.123527}, \href
  {http://adsabs.harvard.edu/abs/2016PhRvD..93l3527C} {93, 123527}

\bibitem[\protect\citeauthoryear{{Dav{\'e}}, {Spergel}, {Steinhardt}  \&
  {Wandelt}}{{Dav{\'e}} et~al.}{2001}]{Dave01}
{Dav{\'e}} R.,  {Spergel} D.~N.,  {Steinhardt} P.~J.,   {Wandelt} B.~D.,  2001,
  \mn@doi [\apj] {10.1086/318417}, \href
  {http://adsabs.harvard.edu/abs/2001ApJ...547..574D} {547, 574}
  
  \bibitem[\protect\citeauthoryear{{Di Cintio}, {Knebe}, {Libeskind}, {Yepes},
  {Gottl{\"o}ber}  \& {Hoffman}}{{Di Cintio} et~al.}{2011}]{diCintio11}
{Di Cintio} A.,  {Knebe} A.,  {Libeskind} N.~I.,  {Yepes} G.,  {Gottl{\"o}ber}
  S.,   {Hoffman} Y.,  2011, \mn@doi [\mnras]
  {10.1111/j.1745-3933.2011.01123.x}, \href
  {http://adsabs.harvard.edu/abs/2011MNRAS.417L..74D} {417, L74}

\bibitem[\protect\citeauthoryear{{Einasto}}{{Einasto}}{1965}]{Einasto65}
{Einasto} J.,  1965, Trudy Astrofizicheskogo Instituta Alma-Ata, \href
  {http://adsabs.harvard.edu/abs/1965TrAlm...5...87E} {5, 87}

\bibitem[\protect\citeauthoryear{{Eisenstein} et~al.,}{{Eisenstein}
  et~al.}{2005}]{Eisenstein05}
{Eisenstein} D.~J.,  et~al., 2005, \mn@doi [\apj] {10.1086/466512}, \href
  {https://ui.adsabs.harvard.edu/abs/2005ApJ...633..560E} {633, 560}
  
 \bibitem[\protect\citeauthoryear{Feng, Kaplinghat, \& Yu}{2010}]{Feng10} Feng J.~L., Kaplinghat M., Yu H.-B., 2010, PhRvL, 104, 151301. doi:10.1103/PhysRevLett.104.151301

\bibitem[\protect\citeauthoryear{{Fitts} et~al.,}{{Fitts}
  et~al.}{2019}]{Fitts19}
{Fitts} A.,  et~al., 2019, \mn@doi [\mnras] {10.1093/mnras/stz2613}, \href
  {https://ui.adsabs.harvard.edu/abs/2019MNRAS.490..962F} {490, 962}

\bibitem[\protect\citeauthoryear{{Gilmore}, {Wilkinson}, {Wyse}, {Kleyna},
  {Koch}, {Evans}  \& {Grebel}}{{Gilmore} et~al.}{2007}]{Gilmore07}
{Gilmore} G.,  {Wilkinson} M.~I.,  {Wyse} R.~F.~G.,  {Kleyna} J.~T.,  {Koch}
  A.,  {Evans} N.~W.,   {Grebel} E.~K.,  2007, \mn@doi [\apj] {10.1086/518025},
  \href {http://adsabs.harvard.edu/abs/2007ApJ...663..948G} {663, 948}

\bibitem[\protect\citeauthoryear{{Glass}}{{Glass}}{2010}]{Glass10}
{Glass} E.~N.,  2010, \mn@doi [\prd] {10.1103/PhysRevD.82.044039}, \href
  {https://ui.adsabs.harvard.edu/abs/2010PhRvD..82d4039G} {82, 044039}

\bibitem[\protect\citeauthoryear{{Gregory}, {Collins}, {Read}, {Irwin},
  {Ibata}, {Martin}, {McConnachie}  \& {Weisz}}{{Gregory}
  et~al.}{2019}]{Gregory19}
{Gregory} A.~L.,  {Collins} M. L.~M.,  {Read} J.~I.,  {Irwin} M.~J.,  {Ibata}
  R.~A.,  {Martin} N.~F.,  {McConnachie} A.~W.,   {Weisz} D.~R.,  2019, \mn@doi
  [\mnras] {10.1093/mnras/stz518}, \href
  {https://ui.adsabs.harvard.edu/abs/2019MNRAS.485.2010G} {485, 2010}

\bibitem[\protect\citeauthoryear{{Hayashi}, {Chiba}  \& {Ishiyama}}{{Hayashi}
  et~al.}{2020}]{Hayashi20}
{Hayashi} K.,  {Chiba} M.,   {Ishiyama} T.,  2020, arXiv e-prints, \href
  {https://ui.adsabs.harvard.edu/abs/2020arXiv200713780H} {p. arXiv:2007.13780}

\bibitem[\protect\citeauthoryear{{Huo}, {Yu}  \& {Zhong}}{{Huo}
  et~al.}{2020}]{Huo19}
{Huo} R.,  {Yu} H.-B.,   {Zhong} Y.-M.,  2020, \mn@doi [\jcap]
  {10.1088/1475-7516/2020/06/051}, \href
  {https://ui.adsabs.harvard.edu/abs/2020JCAP...06..051H} {2020, 051}
  
  
\bibitem[\protect\citeauthoryear{Kahlhoefer et al.}{2019}]{Kahlhoefer19} 
Kahlhoefer F., Kaplinghat M., Slatyer T.~R., Wu C.-L., 2019,  \mn@doi [\jcap]
{10.1088/1475-7516/2019/12/010}, \href
{https://ui.adsabs.harvard.edu/abs/2019JCAP...12..010K/abstract} {2019, 010}


\bibitem[\protect\citeauthoryear{{Kamada}, {Kaplinghat}, {Pace}  \&
  {Yu}}{{Kamada} et~al.}{2017}]{Kamada17}
{Kamada} A.,  {Kaplinghat} M.,  {Pace} A.~B.,   {Yu} H.-B.,  2017, \mn@doi
  [Physical Review Letters] {10.1103/PhysRevLett.119.111102}, \href
  {http://adsabs.harvard.edu/abs/2017PhRvL.119k1102K} {119, 111102}
 
 
  \bibitem[\protect\citeauthoryear{Kaplinghat, Valli, \& Yu}{2019}]{Kaplinghat19} 
  Kaplinghat M., Valli M., Yu H.-B., 2019, \mn@doi [\mnras] 
  {10.1093/mnras/stz2511}, \href 
  {https://ui.adsabs.harvard.edu/abs/2019MNRAS.490..231K/abstract} {490, 231}

\bibitem[\protect\citeauthoryear{{Klypin}, {Kravtsov}, {Bullock}  \&
  {Primack}}{{Klypin} et~al.}{2001}]{Klypin01}
{Klypin} A.,  {Kravtsov} A.~V.,  {Bullock} J.~S.,   {Primack} J.~R.,  2001,
  \mn@doi [\apj] {10.1086/321400}, \href
  {https://ui.adsabs.harvard.edu/abs/2001ApJ...554..903K} {554, 903}

\bibitem[\protect\citeauthoryear{{Loeb} \& {Weiner}}{{Loeb} \&
  {Weiner}}{2011}]{Loeb11}
{Loeb} A.,  {Weiner} N.,  2011, \mn@doi [Physical Review Letters]
  {10.1103/PhysRevLett.106.171302}, \href
  {http://adsabs.harvard.edu/abs/2011PhRvL.106q1302L} {106, 171302}

\bibitem[\protect\citeauthoryear{{Lovell} et~al.,}{{Lovell}
  et~al.}{2012}]{Lovell12}
{Lovell} M.~R.,  et~al., 2012, \mn@doi [\mnras]
  {10.1111/j.1365-2966.2011.20200.x}, \href
  {http://adsabs.harvard.edu/abs/2012MNRAS.420.2318L} {420, 2318}

\bibitem[\protect\citeauthoryear{{Lovell}, {Frenk}, {Eke}, {Jenkins}, {Gao}  \&
  {Theuns}}{{Lovell} et~al.}{2014}]{Lovell14}
{Lovell} M.~R.,  {Frenk} C.~S.,  {Eke} V.~R.,  {Jenkins} A.,  {Gao} L.,
  {Theuns} T.,  2014, \mn@doi [\mnras] {10.1093/mnras/stt2431}, \href
  {http://adsabs.harvard.edu/abs/2014MNRAS.439..300L} {439, 300}

\bibitem[\protect\citeauthoryear{{Lovell} et~al.,}{{Lovell}
  et~al.}{2018}]{Lovell18b}
{Lovell} M.~R.,  et~al., 2018, \mn@doi [\mnras] {10.1093/mnras/sty2339}, \href
  {https://ui.adsabs.harvard.edu/#abs/2018MNRAS.481.1950L} {481, 1950}

\bibitem[\protect\citeauthoryear{Lovell et al.}{2021}]{Lovell21} Lovell M.~R., Cautun M., Frenk C.~S., Hellwing W.~A., Newton O., 2021, arXiv e-prints, \href{https://ui.adsabs.harvard.edu/abs/2021arXiv210403322L}{p.arxiv:2104.03322} 
    

\bibitem[\protect\citeauthoryear{{Lynden-Bell} \& {Wood}}{{Lynden-Bell} \&
  {Wood}}{1968}]{LyndenBell68}
{Lynden-Bell} D.,  {Wood} R.,  1968, \mn@doi [\mnras]
  {10.1093/mnras/138.4.495}, \href
  {https://ui.adsabs.harvard.edu/abs/1968MNRAS.138..495L} {138, 495}

\bibitem[\protect\citeauthoryear{{Moore}, {Governato}, {Quinn}, {Stadel}  \&
  {Lake}}{{Moore} et~al.}{1998}]{Moore98}
{Moore} B.,  {Governato} F.,  {Quinn} T.,  {Stadel} J.,   {Lake} G.,  1998,
  \mn@doi [\apjl] {10.1086/311333}, \href
  {https://ui.adsabs.harvard.edu/abs/1998ApJ...499L...5M} {499, L5}

\bibitem[\protect\citeauthoryear{{Navarro}, {Eke}  \& {Frenk}}{{Navarro}
  et~al.}{1996a}]{NEF96}
{Navarro} J.~F.,  {Eke} V.~R.,   {Frenk} C.~S.,  1996a, \mnras, \href
  {http://adsabs.harvard.edu/abs/1996MNRAS.283L..72N} {283, L72}

\bibitem[\protect\citeauthoryear{{Navarro}, {Frenk}  \& {White}}{{Navarro}
  et~al.}{1996b}]{NFW_96}
{Navarro} J.~F.,  {Frenk} C.~S.,   {White} S.~D.~M.,  1996b, \mn@doi [\apj]
  {10.1086/177173}, \href {http://adsabs.harvard.edu/abs/1996ApJ...462..563N}
  {462, 563}

\bibitem[\protect\citeauthoryear{{Navarro}, {Frenk}  \& {White}}{{Navarro}
  et~al.}{1997}]{NFW_97}
{Navarro} J.~F.,  {Frenk} C.~S.,   {White} S.~D.~M.,  1997, \mn@doi [\apj]
  {10.1086/304888}, \href {http://adsabs.harvard.edu/abs/1997ApJ...490..493N}
  {490, 493}

\bibitem[\protect\citeauthoryear{{Nishikawa}, {Boddy}  \&
  {Kaplinghat}}{{Nishikawa} et~al.}{2020}]{Nishikawa20}
{Nishikawa} H.,  {Boddy} K.~K.,   {Kaplinghat} M.,  2020, \mn@doi [\prd]
  {10.1103/PhysRevD.101.063009}, \href
  {https://ui.adsabs.harvard.edu/abs/2020PhRvD.101f3009N} {101, 063009}

\bibitem[\protect\citeauthoryear{{Oman} et~al.,}{{Oman} et~al.}{2015}]{Oman15}
{Oman} K.~A.,  et~al., 2015, \mn@doi [\mnras] {10.1093/mnras/stv1504}, \href
  {https://ui.adsabs.harvard.edu/abs/2015MNRAS.452.3650O} {452, 3650}

\bibitem[\protect\citeauthoryear{{Oman}, {Marasco}, {Navarro}, {Frenk},
  {Schaye}  \& {Ben{\'\i}tez-Llambay}}{{Oman} et~al.}{2019}]{Oman19}
{Oman} K.~A.,  {Marasco} A.,  {Navarro} J.~F.,  {Frenk} C.~S.,  {Schaye} J.,
  {Ben{\'\i}tez-Llambay} A.~r.,  2019, \mn@doi [\mnras]
  {10.1093/mnras/sty2687}, \href
  {https://ui.adsabs.harvard.edu/abs/2019MNRAS.482..821O} {482, 821}

\bibitem[\protect\citeauthoryear{{Onions} et~al.,}{{Onions}
  et~al.}{2012}]{Onions12}
{Onions} J.,  et~al., 2012, \mn@doi [\mnras]
  {10.1111/j.1365-2966.2012.20947.x}, \href
  {https://ui.adsabs.harvard.edu/abs/2012MNRAS.423.1200O} {423, 1200}

\bibitem[\protect\citeauthoryear{Peter, Rocha, Bullock  \& Kaplinghat}{Peter
  et~al.}{2013}]{Peter13}
Peter A. H.~G.,  Rocha M.,  Bullock J.~S.,   Kaplinghat M.,  2013, \mn@doi
  [Mon. Not. Roy. Astron. Soc.] {10.1093/mnras/sts535}, 430, 105

\bibitem[\protect\citeauthoryear{{Planck Collaboration} et~al.,}{{Planck
  Collaboration} et~al.}{2016}]{Planck16}
{Planck Collaboration} et~al., 2016, \mn@doi [\aap]
  {10.1051/0004-6361/201525830}, \href
  {http://adsabs.harvard.edu/abs/2016A%26A...594A..13P} {594, A13}

\bibitem[\protect\citeauthoryear{{Pontzen} \& {Governato}}{{Pontzen} \&
  {Governato}}{2012}]{Pontzen_Governato_11}
{Pontzen} A.,  {Governato} F.,  2012, \mn@doi [\mnras]
  {10.1111/j.1365-2966.2012.20571.x}, \href
  {http://adsabs.harvard.edu/abs/2012MNRAS.421.3464P} {421, 3464}

\bibitem[\protect\citeauthoryear{{Power}, {Navarro}, {Jenkins}, {Frenk},
  {White}, {Springel}, {Stadel}  \& {Quinn}}{{Power} et~al.}{2003}]{Power03}
{Power} C.,  {Navarro} J.~F.,  {Jenkins} A.,  {Frenk} C.~S.,  {White} S.~D.~M.,
   {Springel} V.,  {Stadel} J.,   {Quinn} T.,  2003, \mn@doi [\mnras]
  {10.1046/j.1365-8711.2003.05925.x}, \href
  {http://adsabs.harvard.edu/abs/2003MNRAS.338...14P} {338, 14}

\bibitem[\protect\citeauthoryear{{Rocha}, {Peter}, {Bullock}, {Kaplinghat},
  {Garrison-Kimmel}, {O{\~n}orbe}  \& {Moustakas}}{{Rocha}
  et~al.}{2013}]{Rocha13}
{Rocha} M.,  {Peter} A.~H.~G.,  {Bullock} J.~S.,  {Kaplinghat} M.,
  {Garrison-Kimmel} S.,  {O{\~n}orbe} J.,   {Moustakas} L.~A.,  2013, \mn@doi
  [\mnras] {10.1093/mnras/sts514}, \href
  {http://adsabs.harvard.edu/abs/2013MNRAS.430...81R} {430, 81}

\bibitem[\protect\citeauthoryear{{Sameie}, {Creasey}, {Yu}, {Sales},
  {Vogelsberger}  \& {Zavala}}{{Sameie} et~al.}{2018}]{Sameie18}
{Sameie} O.,  {Creasey} P.,  {Yu} H.-B.,  {Sales} L.~V.,  {Vogelsberger} M.,
  {Zavala} J.,  2018, \mn@doi [\mnras] {10.1093/mnras/sty1516}, \href
  {https://ui.adsabs.harvard.edu/abs/2018MNRAS.479..359S} {479, 359}

\bibitem[\protect\citeauthoryear{{Sameie}, {Yu}, {Sales}, {Vogelsberger}  \&
  {Zavala}}{{Sameie} et~al.}{2020}]{Sameie20}
{Sameie} O.,  {Yu} H.-B.,  {Sales} L.~V.,  {Vogelsberger} M.,   {Zavala} J.,
  2020, \mn@doi [\prl] {10.1103/PhysRevLett.124.141102}, \href
  {https://ui.adsabs.harvard.edu/abs/2020PhRvL.124n1102S} {124, 141102}

\bibitem[\protect\citeauthoryear{{Santos-Santos} et~al.,}{{Santos-Santos}
  et~al.}{2020}]{Santos19}
{Santos-Santos} I. M.~E.,  et~al., 2020, \mn@doi [\mnras]
  {10.1093/mnras/staa1072}, \href
  {https://ui.adsabs.harvard.edu/abs/2020MNRAS.495...58S} {495, 58}

\bibitem[\protect\citeauthoryear{{Sawala} et~al.,}{{Sawala}
  et~al.}{2016}]{Sawala16a}
{Sawala} T.,  et~al., 2016, \mn@doi [\mnras] {10.1093/mnras/stw145}, \href
  {http://adsabs.harvard.edu/abs/2016MNRAS.457.1931S} {457, 1931}

\bibitem[\protect\citeauthoryear{{Schewtschenko}, {Baugh}, {Wilkinson},
  {B{\oe}hm}, {Pascoli}  \& {Sawala}}{{Schewtschenko}
  et~al.}{2016}]{Schewtschenko16}
{Schewtschenko} J.~A.,  {Baugh} C.~M.,  {Wilkinson} R.~J.,  {B{\oe}hm} C.,
  {Pascoli} S.,   {Sawala} T.,  2016, \mn@doi [\mnras] {10.1093/mnras/stw1078},
  \href {https://ui.adsabs.harvard.edu/abs/2016MNRAS.461.2282S} {461, 2282}

\bibitem[\protect\citeauthoryear{{Spergel} \& {Steinhardt}}{{Spergel} \&
  {Steinhardt}}{2000}]{Spergel00}
{Spergel} D.~N.,  {Steinhardt} P.~J.,  2000, \mn@doi [Physical Review Letters]
  {10.1103/PhysRevLett.84.3760}, \href
  {http://adsabs.harvard.edu/abs/2000PhRvL..84.3760S} {84, 3760}

\bibitem[\protect\citeauthoryear{{Spergel} et~al.,}{{Spergel}
  et~al.}{2003}]{wmap1}
{Spergel} D.~N.,  et~al., 2003, \mn@doi [\apjs] {10.1086/377226}, \href
  {http://adsabs.harvard.edu/abs/2003ApJS..148..175S} {148, 175}

\bibitem[\protect\citeauthoryear{{Springel}, {White}, {Tormen}  \&
  {Kauffmann}}{{Springel} et~al.}{2001}]{Springel01}
{Springel} V.,  {White} S.~D.~M.,  {Tormen} G.,   {Kauffmann} G.,  2001,
  \mn@doi [\mnras] {10.1046/j.1365-8711.2001.04912.x}, \href
  {http://adsabs.harvard.edu/abs/2001MNRAS.328..726S} {328, 726}

\bibitem[\protect\citeauthoryear{{Springel} et~al.,}{{Springel}
  et~al.}{2008}]{Springel08b}
{Springel} V.,  et~al., 2008, \mnras, \href
  {http://adsabs.harvard.edu/abs/2008MNRAS.391.1685S} {391, 1685}

\bibitem[\protect\citeauthoryear{{Strigari}, {Frenk}  \& {White}}{{Strigari}
  et~al.}{2010}]{Strigari10}
{Strigari} L.~E.,  {Frenk} C.~S.,   {White} S.~D.~M.,  2010, \mn@doi [\mnras]
  {10.1111/j.1365-2966.2010.17287.x}, \href
  {http://adsabs.harvard.edu/abs/2010MNRAS.408.2364S} {408, 2364}

\bibitem[\protect\citeauthoryear{{Strigari}, {Frenk}  \& {White}}{{Strigari}
  et~al.}{2017}]{Strigari17}
{Strigari} L.~E.,  {Frenk} C.~S.,   {White} S. D.~M.,  2017, \mn@doi [\apj]
  {10.3847/1538-4357/aa5c8e}, \href
  {https://ui.adsabs.harvard.edu/abs/2017ApJ...838..123S} {838, 123}

\bibitem[\protect\citeauthoryear{{Vogelsberger}, {Zavala}  \&
  {Loeb}}{{Vogelsberger} et~al.}{2012}]{Vogelsberger12}
{Vogelsberger} M.,  {Zavala} J.,   {Loeb} A.,  2012, \mn@doi [\mnras]
  {10.1111/j.1365-2966.2012.21182.x}, \href
  {https://ui.adsabs.harvard.edu/\#abs/2012MNRAS.423.3740V} {423, 3740}

\bibitem[\protect\citeauthoryear{{Vogelsberger}, {Zavala}, {Cyr-Racine},
  {Pfrommer}, {Bringmann}  \& {Sigurdson}}{{Vogelsberger}
  et~al.}{2016}]{Vogelsberger16}
{Vogelsberger} M.,  {Zavala} J.,  {Cyr-Racine} F.-Y.,  {Pfrommer} C.,
  {Bringmann} T.,   {Sigurdson} K.,  2016, \mn@doi [\mnras]
  {10.1093/mnras/stw1076}, \href
  {http://adsabs.harvard.edu/abs/2016MNRAS.460.1399V} {460, 1399}

\bibitem[\protect\citeauthoryear{{Vogelsberger}, {Zavala}, {Schutz}  \&
  {Slatyer}}{{Vogelsberger} et~al.}{2019}]{Vogelsberger2019}
{Vogelsberger} M.,  {Zavala} J.,  {Schutz} K.,   {Slatyer} T.~R.,  2019,
  \mn@doi [\mnras] {10.1093/mnras/stz340}, \href
  {https://ui.adsabs.harvard.edu/\#abs/2019MNRAS.tmp..346V} {p.~346}

\bibitem[\protect\citeauthoryear{{Walker} \& {Pe{\~n}arrubia}}{{Walker} \&
  {Pe{\~n}arrubia}}{2011}]{Walker11}
{Walker} M.~G.,  {Pe{\~n}arrubia} J.,  2011, \mn@doi [\apj]
  {10.1088/0004-637X/742/1/20}, \href
  {http://adsabs.harvard.edu/abs/2011ApJ...742...20W} {742, 20}

\bibitem[\protect\citeauthoryear{{Walker}, {Mateo}, {Olszewski},
  {Pe{\~n}arrubia}, {Wyn Evans}  \& {Gilmore}}{{Walker}
  et~al.}{2009}]{Walker09}
{Walker} M.~G.,  {Mateo} M.,  {Olszewski} E.~W.,  {Pe{\~n}arrubia} J.,  {Wyn
  Evans} N.,   {Gilmore} G.,  2009, \mn@doi [\apj]
  {10.1088/0004-637X/704/2/1274}, \href
  {http://adsabs.harvard.edu/abs/2009ApJ...704.1274W} {704, 1274}

\bibitem[\protect\citeauthoryear{{Wolf}, {Martinez}, {Bullock}, {Kaplinghat},
  {Geha}, {Mu{\~n}oz}, {Simon}  \& {Avedo}}{{Wolf} et~al.}{2010}]{Wolf10}
{Wolf} J.,  {Martinez} G.~D.,  {Bullock} J.~S.,  {Kaplinghat} M.,  {Geha} M.,
  {Mu{\~n}oz} R.~R.,  {Simon} J.~D.,   {Avedo} F.~F.,  2010, \mn@doi [\mnras]
  {10.1111/j.1365-2966.2010.16753.x}, \href
  {http://adsabs.harvard.edu/abs/2010MNRAS.406.1220W} {406, 1220}

\bibitem[\protect\citeauthoryear{{Zavala}, {Vogelsberger}  \&
  {Walker}}{{Zavala} et~al.}{2013}]{Zavala13}
{Zavala} J.,  {Vogelsberger} M.,   {Walker} M.~G.,  2013, \mn@doi [\mnras]
  {10.1093/mnrasl/sls053}, \href
  {http://adsabs.harvard.edu/abs/2013MNRAS.431L..20Z} {431, L20}

\bibitem[\protect\citeauthoryear{{Zavala}, {Lovell}, {Vogelsberger}  \&
  {Burger}}{{Zavala} et~al.}{2019}]{Zavala2019}
{Zavala} J.,  {Lovell} M.~R.,  {Vogelsberger} M.,   {Burger} J.~D.,  2019,
  \mn@doi [\prd] {10.1103/PhysRevD.100.063007}, \href
  {https://ui.adsabs.harvard.edu/abs/2019PhRvD.100f3007Z} {100, 063007}


\makeatother
\end{thebibliography}


\bsp	
\label{lastpage}
\end{document}